\documentclass[letterpaper,titlepage,12pt]{article}
\usepackage[colorlinks=true,urlcolor=blue,citecolor=magenta]{hyperref}
\usepackage{amssymb,amsmath,amsfonts}
\usepackage{epsfig}
\usepackage{epsfig}
\usepackage{graphicx}
\usepackage{epstopdf}
\usepackage{caption}
\usepackage{subcaption}
\usepackage{amscd}
\usepackage{amsthm}
\usepackage{latexsym}
\usepackage{amsbsy}
\usepackage{bbm}
\usepackage[english]{babel}
\usepackage{psfrag}
\usepackage{tabularx}

\allowdisplaybreaks

\def\clock{{\count0=\time
           \divide\count0 60
           \ifnum\count0<10 0\fi\the\count0
           \multiply\count0 -60 \advance\count0 \time
           :\ifnum\count0<10 0\fi \the\count0
         }}
\newcommand{\timestamp}{{\small\vbox{\hbox{\tt\jobname.tex}
\hbox{\the\day/\the\month/\the\year, \clock}}}}

\numberwithin{equation}{section}

\begin{document}

\begin{titlepage}
\rightline{\vbox{   \phantom{ghost} }}
%
%
 \vskip 1.4 cm
\centerline{\LARGE \bf Gauging the Carroll Algebra and}
\vspace{.2cm}
\centerline{\LARGE \bf Ultra-Relativistic Gravity}

\vskip 1.5cm

\centerline{\large {{\bf Jelle Hartong}}}

\vskip .8cm

\begin{center}

\sl Physique Th\'eorique et Math\'ematique and International Solvay Institutes,\\
Universit\'e Libre de Bruxelles, C.P. 231, 1050 Brussels, Belgium.
\vskip 0.4cm

\end{center}
\vskip 0.6cm


\vskip .8cm \centerline{\bf Abstract} \vskip 0.2cm \noindent

It is well known that the geometrical framework of Riemannian geometry that underlies general relativity and its torsionful extension to Riemann--Cartan geometry can be obtained from a procedure known as gauging the Poincar\'e algebra. Recently it has been shown that gauging the centrally extended Galilei algebra, known as the Bargmann algebra, leads to  a geometrical framework that when made dynamical gives rise to Ho\v rava--Lifshitz gravity. Here we consider the case where we contract the Poincar\'e algebra by sending the speed of light to zero leading to the Carroll algebra. We show how this algebra can be gauged and we construct the most general affine connection leading to the geometry of so-called Carrollian space-times. Carrollian space-times appear for example as the geometry on null hypersurfaces in a Lorentzian space-time of one dimension higher. We also construct theories of  ultra-relativistic (Carrollian) gravity in 2+1 dimensions with dynamical exponent $z<1$ including cases that have anisotropic Weyl invariance for $z=0$.

\end{titlepage}

\pagestyle{empty}
\small
\tableofcontents

\normalsize
\newpage
\pagestyle{plain}
\setcounter{page}{1}


\section{Introduction}

Over the recent years it has become clear that non-relativistic symmetry groups play an important role in many examples of non-AdS holography. This has been made most apparent in the case of Lifshitz holography where it is has been shown that the boundary geometry is described by Newton--Cartan geometry in the presence of torsion \cite{Hartong:2014oma,Bergshoeff:2014uea,Hartong:2015wxa}. Further also in the case of Schr\"odinger holography there are many hints that the boundary field theory couples to a certain non-Riemannian geometry \cite{Guica:2010sw,Hartong:2010ec,Hartong:2013cba,Andrade:2014iia,Andrade:2014kba}. In AdS/CFT the fact that the boundary geometry is described by Riemannian geometry, just like the bulk geometry, is a special feature of the precise fall-off of the AdS metric (and its asymptotically locally AdS generalizations \cite{deHaro:2000xn,Papadimitriou:2005ii}) near the boundary. It is however not expected that the Riemannian nature of the boundary geometry seen in AdS/CFT is a generic feature in other non-AdS holographic dualities. Hence in order to better understand known candidates for non-AdS holography we must learn how to describe various non-Riemannian geometries. 

Recently it has been argued that the Carroll algebra, which can be obtained as an ultra-relativistic limit ($c\rightarrow 0$) of the Poincar\'e algebra \cite{LŽvy1965,Bacry:1968zf}, plays an important role in flat space holography \cite{Duval:2014uva}. The $c\rightarrow 0$ contraction of the Poincar\'e algebra results in a peculiar light cone structure where the light cone has collapsed to a line. The Carroll algebra is given by
\begin{eqnarray*}
\left[J_{ab},P_c\right]&=&\delta_{ac}P_b-\delta_{bc}P_a\,,\\
\left[J_{ab},C_c\right]&=&\delta_{ac}C_b-\delta_{bc}C_a\,,\\
\left[J_{ab},J_{cd}\right]&=&\delta_{ac}J_{bd}-\delta_{ad}J_{bc}-\delta_{bc}J_{ad}+\delta_{bd}J_{ac}\,,\\
\left[P_a,C_b\right]&=&\delta_{ab}H\,,
\end{eqnarray*}
where $a=1,\ldots,d$. In here $H$ is the Hamiltonian, $P_a$ spatial momenta, $J_{ab}$ spatial rotations and $C_a$ Carrollian boosts. In Cartesian coordinates with time $t$ and space coordinates $x^i$ a Carrollian boost means $t\rightarrow t+\vec v\cdot\vec x$. In \cite{Duval:2014uva} it is shown that future and past null infinity form Carrollian space-times and that the BMS algebra forms a conformal extension of the Carroll algebra. It is therefore of interest to understand in full generality what Carrollian space-times are and how field theories couple to them (see e.g. the work of \cite{Hofman:2014loa} on coupling warped conformal field theories to geometries obtained by gauging the Carroll algebra). When gauging this algebra we associate vielbeins $\tau_\mu$ and $e_\mu^a$ to the time and space translation generators $H$ and $P_a$, respectively.

For the case of relativistic field theories we know that they couple to Riemannian geometry. The latter and its torsionful extension, known as Riemann--Cartan geometry, can be obtained by a procedure known as gauging the Poincar\'e algebra (see e.g. appendix A of \cite{Hartong:2015zia}). Similar gauging techniques also allow one to describe torsional Newton--Cartan (TNC) geometry which was found in a holographic context in \cite{Christensen:2013lma,Christensen:2013rfa,Hartong:2014oma,Hartong:2015wxa}. We refer to \cite{Son:2013rqa,Geracie:2014nka,Banerjee:2014nja,Brauner:2014jaa,Jensen:2014aia,Hartong:2014pma,Hartong:2015wxa} for the use of TNC geometry in a field theoretical context (see \cite{Bekaert:2014bwa} for a nice geometrical account of TNC geometry). In order to obtain torsional Newton--Cartan geometry one gauges the centrally extended Galilei algebra known as the Bargmann algebra \cite{Andringa:2010it,Bergshoeff:2014uea,Hartong:2015zia}. 

Given the relevance of the Carroll algebra to flat space holography it is a natural question to ask if we can gauge the Carroll algebra and what the resulting geometrical structure is. The gauging of the Carroll algebra will be discussed in section \ref{sec:gaugingCarroll}. First this is done in full generality involving the Carrollian vielbeins $\tau_\mu$, $e_\mu^a$ and a Carrollian metric compatible affine connection $\Gamma^\rho_{\mu\nu}$. Then we introduce a contravariant vector $M^\mu$ and show that the Carrollian metric compatible affine connection $\Gamma^\rho_{\mu\nu}$ can fully be written in terms of $\tau_\mu$, $e_\mu^a$ and $M^\mu$. The role of $M^\mu$ is to ensure that $\Gamma^\rho_{\mu\nu}$ when written in terms of the Carrollian vielbeins remains invariant under local (tangent space) Carrollian boosts. In the next section, section \ref{sec:nullhypersurfaces}, it will be shown that the resulting geometrical structure can be realized as the geometry induced on a null hypersurface embedded in a Lorentzian space-time of one dimension higher. Further in section \ref{sec:nullhypersurfaces} we show that the duality between Newton--Cartan and Carrollian space-times observed in \cite{Duval:2014uoa} can be extended when we include the vector $M^\mu$. The Newton--Cartan dual of $M^\mu$ is a co-vector $M_\mu$ that can be written as $m_\mu-\partial_\mu\chi$ where $m_\mu$ is the connection corresponding to the Bargmann extension of the Galilei algebra and $\chi$ is a St\"uckelberg scalar that must be added to the formalism whenever there is torsion \cite{Hartong:2014oma,Bergshoeff:2014uea,Hartong:2015zia}.

In \cite{Hartong:2015zia} it has been shown that when torsional Newton--Cartan geometry is made dynamical the resulting theory of gravity corresponds to Ho\v rava--Lifshitz gravity \cite{Horava:2008ih,Horava:2009uw} including the extension of \cite{Horava:2010zj}. More specifically, depending on the type of torsion, one is either dealing with projectable (no torsion) or non-projectable (so-called twistless torsion) HL gravity. In both these cases there is a preferred foliation of spacelike hypersurfaces. The case of general torsion is an extension of HL gravity in which the timelike vielbein is not required to be hypersurface orthogonal. Since the tangent space group in HL gravity is the Galilean group it is natural to refer to this type of gravitational theories as non-relativistic theories of gravity. In the same spirit, the Carrollian geometry can be made dynamical. We do this using an effective action approach. By this we mean that we assign dilatation weights to the Carrollian fields $\tau_\mu$, $e_\mu^a$ and $M^\mu$ and allow for all possible terms that are relevant or marginal and invariant under local tangent space (Carrollian) transformations. Since the tangent space light cone structure is ultra-relativistic we refer to this as ultra-relativistic gravity. They naturally come with a dynamical exponent $z<1$. We show that for $z=0$ one can construct actions that are invariant under anisotropic Weyl rescalings of the Carrollian fields $\tau_\mu$, $e_\mu^a$ and $M^\mu$. All of this will be the subject of section \ref{sec:Carrollgrav}. 

A special case of such Carrollian theories of gravity are obtained from the ultralocal (in the sense of no space derivatives) limit of general relativity (GR) that was studied in \cite{Teitelboim:1981fb,Henneaux:1979vn} by sending the speed of light to zero. This Carrollian limit of GR is also referred to as the strong coupling limit in which Newton's constant tends to infinity as this has the same effect as sending $c$ to zero \cite{Dautcourt:1997hb,Anderson:2002zn}. Further the Carrollian limit also features in tachyon condensation \cite{Gibbons:2002tv} and cosmological billiards \cite{Damour:2002et}.

{\bf Note added:}
While this manuscript was being finalized, the preprint \cite{Bekaert:2015xua} appeared on the arXiv, which overlaps with the results of sections \ref{sec:gaugingCarroll} and \ref{sec:nullhypersurfaces}.

\section{Gauging the Carroll Algebra}\label{sec:gaugingCarroll}

\subsection{From local Carroll to diffeomorphisms and Carrollian light cones}

The Carroll algebra is obtained as a contraction of the Poincar\'e algebra by sending the speed of light to zero \cite{LŽvy1965,Bacry:1968zf}. The nonzero commutators of the Carroll algebra are
\begin{eqnarray}
\left[J_{ab},P_c\right]&=&\delta_{ac}P_b-\delta_{bc}P_a\,,\\
\left[J_{ab},C_c\right]&=&\delta_{ac}C_b-\delta_{bc}C_a\,,\\
\left[J_{ab},J_{cd}\right]&=&\delta_{ac}J_{bd}-\delta_{ad}J_{bc}-\delta_{bc}J_{ad}+\delta_{bd}J_{ac}\,,\\
\left[P_a,C_b\right]&=&\delta_{ab}H\,,
\end{eqnarray}
where $a=1,\ldots,d$. We thus see that the algebra is isomorphic to the semi-direct product of $SO(d)$ with the Heisenberg algebra whose central element is the Hamiltonian. 

In order to gauge the algebra we follow the procedure of \cite{Andringa:2010it,Hartong:2015zia} where the gauging of the Galilei algebra and its central extension, the Bargmann algebra, were discussed (without torsion \cite{Andringa:2010it} and including torsion \cite{Hartong:2015zia}). For an earlier discussion of gauging the Carroll algebra see  \cite{Bergshoeff:2014jla}.

We define a connection $\mathcal{A}_\mu$ as
\begin{equation}
\mathcal{A}_\mu=H\tau_\mu+P_a e^a_\mu+C_a\Omega_\mu{}^a+\frac{1}{2}J_{ab}\Omega_\mu{}^{ab}\,,
\end{equation}
where $\mu$ takes $d+1$ values related to the fact that there is one time and $d$ space translation generators. We thus work with a $(d+1)$-dimensional space-time. This connection transforms in the adjoint as
\begin{equation}
\delta\mathcal{A}_\mu=\partial_\mu\Lambda+[A_\mu,\Lambda]\,.
\end{equation}
Without loss of generality we can write $\Lambda$ as
\begin{equation}
\Lambda=\xi^\mu\mathcal{A}_\mu+\Sigma\,,
\end{equation}
where $\Sigma$ is given by 
\begin{equation}
\Sigma=C_a\lambda^a+\frac{1}{2}J_{ab}\lambda^{ab}\,.
\end{equation}
We would like to think of $\xi^\mu$ as the generator of diffeomorphisms and $\Sigma$ as the internal (tangent) space transformations. To this end we introduce a new local transformation denoted by $\bar\delta$ that is defined as
\begin{equation}
\bar\delta\mathcal{A}_\mu=\delta\mathcal{A}_\mu-\xi^\nu\mathcal{F}_{\mu\nu}=\mathcal{L}_\xi\mathcal{A}_\mu+\partial_\mu\Sigma+[A_\mu,\Sigma]\,,
\end{equation}
where $\mathcal{F}_{\mu\nu}$ is the field strength
\begin{eqnarray}
\mathcal{F}_{\mu\nu} & = & \partial_\mu\mathcal{A}_\nu-\partial_\nu\mathcal{A}_\mu+[\mathcal{A}_\mu,\mathcal{A}_\nu]\nonumber\\
& = & H R_{\mu\nu}(H)+P_a R_{\mu\nu}{}^a(P)+C_a R_{\mu\nu}{}^a(C)+\frac{1}{2}J_{ab}R_{\mu\nu}{}^{ab}(J)\,.
\end{eqnarray}
In components the $\bar\delta$ transformations act as
\begin{eqnarray}
\bar\delta\tau_\mu & = & \mathcal{L}_\xi\tau_\mu+e^a_\mu\lambda_a\,,\\
\bar\delta e_\mu^a & = & \mathcal{L}_\xi e_\mu^a+\lambda^a{}_b e^b_\mu\,,\\
\bar\delta\Omega_\mu{}^a & = & \mathcal{L}_\xi\Omega_\mu{}^a+\partial_\mu\lambda^a+\lambda^a{}_b\Omega_\mu{}^b-\lambda_b\Omega_\mu{}^{ab}\,,\\
\bar\delta\Omega_\mu{}^{ab} & = & \mathcal{L}_\xi\Omega_\mu{}^{ab}+\partial_\mu\lambda^{ab}+\lambda^a{}_c\Omega_\mu{}^{cb}-\lambda^b{}_c\Omega_\mu{}^{ca}\,.
\end{eqnarray}
The Lie derivatives along $\xi^\mu$ correspond to the generators of general coordinate transformations whereas the remaining local transformations with parameters $\lambda^a$ and $\lambda^{ab}$ correspond to local tangent space transformations\footnote{We emphasize that in order to obtain the $\bar\delta$ transformations, i.e. the diffeomorphisms and local tangent space transformations we did not need to impose any so-called curvature constraints. For a discussion of the curvature constraints we refer the reader to section \ref{subsec:vectorM}.}. The tangent space has a Carrollian light cone structure by which we mean that the light cones have collapsed to a line. This can be seen from the fact that there are no boost transformations acting on the spacelike vielbeins $e_\mu^a$. The component expressions for the field strengths read
\begin{eqnarray}
R_{\mu\nu}(H) & = & \partial_\mu\tau_\nu-\partial_\nu\tau_\mu+e_\mu^a\Omega_{\nu a}-e^a_\nu\Omega_{\mu a}\,,\\
R_{\mu\nu}{}^a(P) & = & \partial_\mu e_\nu^a-\partial_\nu e_\mu^a-\Omega_\mu{}^{ab}e_{\nu b}+\Omega_\nu{}^{ab} e_{\mu b}\,,\\
R_{\mu\nu}{}^a(C) & = & \partial_\mu\Omega_\nu{}^a-\partial_\nu\Omega_\mu{}^a-\Omega_\mu{}^{ab}\Omega_{\nu b}+\Omega_\nu{}^{ab} e_{\mu b}\,,\\
R_{\mu\nu}{}^{ab}(J) & = & \partial_\mu\Omega_\nu{}^{ab}-\partial_\nu\Omega_\mu{}^{ab}-\Omega_\mu{}^{ca}\Omega_\nu{}^b{}_c+\Omega_\nu{}^{ca}\Omega_\mu{}^b{}_c\,.
\end{eqnarray}

\subsection{The affine connection}

The next step is to impose vielbein postulates allowing us to describe the properties of the curvatures in $\mathcal{F}_{\mu\nu}$ in terms of the curvature and torsion of an affine connection $\Gamma^\rho_{\mu\nu}$ that by definition is invariant under the tangent space $\Sigma$ transformations. We define the $\bar\delta$ covariant derivative $\mathcal{D}_\mu$ as
\begin{eqnarray}
\mathcal{D}_\mu\tau_\nu & = & \partial_\mu\tau_\nu-\Gamma^\rho_{\mu\nu}\tau_\rho-\Omega_{\mu a}e^a_\nu\,,\label{eq:covdertau}\\
\mathcal{D}_\mu e^a_\nu & = & \partial_\mu e_\nu^a-\Gamma_{\mu\nu}^\rho e^a_\rho-\Omega_\mu{}^a{}_b e^b_\nu\,.
\end{eqnarray}
The form of the covariant derivatives is uniquely fixed by demanding covariance. The vielbein postulates are then 
\begin{eqnarray}
\mathcal{D}_\mu\tau_\nu & = & 0\,,\label{eq:VP1}\\
\mathcal{D}_\mu e_\mu^a & = & 0\,.\label{eq:VP2}
\end{eqnarray}
We choose the right hand side to be zero because i). it obviously transforms covariantly and ii). even if we could write something else that transforms covariantly we can absorb this into the definition of $\Gamma^\rho_{\mu\nu}$. We can now solve for $\Omega_{\mu a}$ and $\Omega_\mu{}^a{}_b$ in terms of $\Gamma^\rho_{\mu\nu}$ by contracting the vielbein postulates with the inverse vielbeins $v^\mu$ and $e^\mu_a$ that are defined via
\begin{equation}
v^\mu\tau_\mu=-1\,,\qquad v^\mu e^a_\mu=0\,,\qquad e^\mu_a\tau_\mu=0\,,\qquad e^\mu_ae^b_\mu=\delta^b_a\,.
\end{equation}
They transform under the $\bar\delta$ transformations as
\begin{eqnarray}
\bar\delta v^\mu & = & \mathcal{L}_\xi v^\mu\,,\\
\bar\delta e^\mu_a & = & \mathcal{L}_\xi e^\mu_a+v^\mu\lambda_a+\lambda_a{}^b e^\mu_b\,,
\end{eqnarray}
and they satisfy the inverse vielbein postulates
\begin{eqnarray}
\mathcal{D}_\mu v^\nu & = & \partial_\mu v^\nu+\Gamma_{\mu\rho}^\nu v^\rho=0\,,\label{eq:VP3}\\
\mathcal{D}_\mu e^\nu_a & = & \partial_\mu e^\nu_a+\Gamma^\nu_{\mu\rho}e^\rho_a-v^\nu\Omega_{\mu a}-\Omega_{\mu a}{}^b e^\nu_b=0\,.\label{eq:VP4}
\end{eqnarray}

From \eqref{eq:covdertau}--\eqref{eq:VP2} it follows that the torsion $\Gamma^\rho_{\mu\nu}$ relates to the curvatures $R_{\mu\nu}(H)$ and $R_{\mu\nu}{}^a(P)$ via
\begin{equation}
2\Gamma^\rho_{[\mu\nu]}=-v^\rho R_{\mu\nu}(H)+e^\rho_a R_{\mu\nu}{}^a(P)\,.
\end{equation}
We can define a Riemann tensor in the usual way as follows
\begin{equation}\label{eq:Riemann}
\left[\nabla_\mu,\nabla_\nu\right]X_\sigma=R_{\mu\nu\sigma}{}^\rho X_\rho-2\Gamma^\rho_{[\mu\nu]}\nabla_\rho X_\sigma\,,
\end{equation}
where $\nabla_\mu$ only contains the affine connection and where $R_{\mu\nu\sigma}{}^\rho$ is given by
\begin{equation}
R_{\mu\nu\sigma}{}^\rho=-\partial_\mu\Gamma^\rho_{\nu\sigma}+\partial_\nu\Gamma^\rho_{\mu\sigma}-\Gamma^\rho_{\mu\lambda}\Gamma^\lambda_{\nu\sigma}+\Gamma^\rho_{\nu\lambda}\Gamma^\lambda_{\mu\sigma}\,.
\end{equation}
Using the vielbein postulates, i.e. the relation between the affine connection and the tangent space connections, we find
\begin{equation}
R_{\mu\nu\sigma}{}^\rho=-v^\rho e_{\sigma a}R_{\mu\nu}{}^a(C)-e_{\sigma a}e^\rho_b R_{\mu\nu}{}^{ab}(J)\,.
\end{equation}

We have traded the connections $\Omega_\mu{}^a$ and $\Omega_\mu{}^{ab}$ for $\Gamma^\rho_{\mu\nu}$. The latter connection has more components and so they cannot all be free. In fact the vielbein postulates \eqref{eq:VP1}--\eqref{eq:VP4} constrain $\Gamma^\rho_{\mu\nu}$ in the following way
\begin{equation}
\nabla_\mu v^\nu=0\,,\qquad \nabla_\mu h_{\nu\rho}=0\,,
\end{equation}
where we defined $h_{\mu\nu}=\delta_{ab}e^a_\mu e^b_\nu$. We will also adopt the notation $h^{\mu\nu}=\delta^{ab}e_a^\mu e_b^\nu$. In order to find out what the independent components of $\Gamma^\rho_{\mu\nu}$ are we will obtain the most general solution to these metric compatibility equations. We note that both $v^\mu$ and $h_{\mu\nu}$ are invariant under the tangent space transformations. They form the notion of Carrollian metrics. 

We start with the condition $\nabla_\mu h_{\nu\rho}=0$. By permuting the indices and summing the resulting equations appropriately we obtain
\begin{equation}\label{eq:symmetricpartGamma}
2\Gamma^\sigma_{(\mu\rho)}h_{\nu\sigma} = \partial_\mu h_{\nu\rho}+\partial_\rho h_{\mu\nu}-\partial_\nu h_{\rho\mu}-2\Gamma^\sigma_{[\mu\nu]}h_{\sigma\rho}-2\Gamma^\sigma_{[\rho\nu]}h_{\mu\sigma}\,.
\end{equation}
Contracting this equation with $v^\nu$ we find
\begin{equation}\label{eq:condition1}
K_{\mu\rho}=-v^\nu h_{\sigma\rho}\Gamma^\sigma_{[\nu\mu]}-v^\nu h_{\sigma\mu}\Gamma^\sigma_{[\nu\rho]}\,,
\end{equation}
where the extrinsic curvature $K_{\mu\rho}$ is defined as
\begin{equation}\label{eq:extrinsiccurv}
K_{\mu\rho}=-\frac{1}{2}\mathcal{L}_v h_{\mu\rho}\,.
\end{equation}
From \eqref{eq:condition1} we conclude that
\begin{equation}\label{eq:X}
\Gamma^\rho_{[\nu\mu]}=\tau_{[\nu}K_{\mu]\lambda}h^{\sigma\lambda}+X^\sigma_{[\nu\mu]}\,,
\end{equation}
where $X^\sigma_{[\nu\mu]}$ is such that 
\begin{equation}\label{eq:propX}
v^\nu h_{\sigma\rho}X^\sigma_{[\nu\mu]}+v^\nu h_{\sigma\mu}X^\sigma_{[\nu\rho]}=0\,.
\end{equation}
Substituting \eqref{eq:X} into \eqref{eq:symmetricpartGamma} and adding $2\Gamma^\sigma_{[\mu\rho]}h_{\nu\sigma}$ to both sides (using \eqref{eq:X}) we obtain
\begin{eqnarray}
2\Gamma^\sigma_{\mu\rho}h_{\nu\sigma} & = & \partial_\mu h_{\nu\rho}+\partial_\rho h_{\mu\nu}-\partial_\nu h_{\rho\mu}+2\tau_{[\nu}K_{\mu]\rho}+2\tau_{[\nu}K_{\rho]\mu}+2\tau_{[\mu}K_{\rho]\nu}\nonumber\\
&&+2X^\sigma_{[\nu\mu]}h_{\sigma\rho}+2X^\sigma_{[\nu\rho]}h_{\sigma\mu}+2X^\sigma_{[\mu\rho]}h_{\nu\sigma}\,.
\end{eqnarray}
Contracting this with $h^{\nu\lambda}$ and using $h_{\nu\sigma}h^{\nu\lambda}=\delta^\lambda_\sigma+\tau_\sigma v^\lambda$ we find the following most general solution to $\nabla_\mu h_{\nu\rho}=0$
\begin{eqnarray}
\Gamma^\lambda_{\mu\rho} & = & -v^\lambda \tau_\sigma\Gamma^\sigma_{\mu\rho}+\frac{1}{2}h^{\nu\lambda}\left(\partial_\mu h_{\nu\rho}+\partial_\rho h_{\mu\nu}-\partial_\nu h_{\rho\mu}\right)-h^{\nu\lambda}\tau_\rho K_{\mu\nu}\nonumber\\
&&+h^{\nu\lambda}\left(X^\sigma_{[\nu\mu]}h_{\sigma\rho}+X^\sigma_{[\nu\rho]}h_{\sigma\mu}+X^\sigma_{[\mu\rho]}h_{\nu\sigma}\right)\,.
\end{eqnarray}
By contracting this result with $v^\rho$ it can be shown that
\begin{equation}
\left(\delta^\lambda_\rho+v^\lambda\tau_\rho\right)\nabla_\mu v^\rho=0\,,
\end{equation}
so that in order to find the most general $\Gamma^\rho_{\mu\nu}$ obeying both $\nabla_\mu v^\nu=0$ and $\nabla_\mu h_{\nu\rho}=0$ it remains to impose
\begin{equation}
\tau_\rho\nabla_\mu v^\rho=0\,.
\end{equation}
This latter condition is equivalent to $v^\rho\nabla_\mu\tau_\rho=0$ so that
\begin{equation}
\Gamma^\sigma_{\mu\rho}\tau_\sigma=\partial_\mu\tau_\rho+X_{\mu\rho}\,,
\end{equation}
with 
\begin{equation}\label{eq:XandboostOmega}
X_{\mu\rho}=-\nabla_\mu\tau_\rho=-\Omega_{\mu a}e^a_\rho\,,
\end{equation}
satisfying
\begin{equation}
v^\rho X_{\mu\rho}=0\,.
\end{equation}

We thus conclude that the most general $\Gamma^\lambda_{\mu\rho}$ is of the form
\begin{eqnarray}
\Gamma^\lambda_{\mu\rho} & = & -v^\lambda\partial_\mu\tau_\rho+\frac{1}{2}h^{\nu\lambda}\left(\partial_\mu h_{\nu\rho}+\partial_\rho h_{\mu\nu}-\partial_\nu h_{\rho\mu}\right)-h^{\nu\lambda}\tau_\rho K_{\mu\nu}\nonumber\\
&&-v^\lambda X_{\mu\rho}+\frac{1}{2}h^{\nu\lambda}Y_{\nu\mu\rho}\,,\label{eq:finalGamma}
\end{eqnarray}\label{eq:Y}
where $Y_{\nu\mu\rho}$ is given by
\begin{equation}
Y_{\nu\mu\rho}=2X^\sigma_{[\nu\mu]}h_{\sigma\rho}+2X^\sigma_{[\nu\rho]}h_{\sigma\mu}+2X^\sigma_{[\mu\rho]}h_{\nu\sigma}\,,
\end{equation}
which has the property that $v^\nu Y_{\nu\mu\rho}=v^\rho Y_{\nu\mu\rho}=0$ as follows from \eqref{eq:propX}. The connection \eqref{eq:finalGamma} has torsion that is given by
\begin{equation}\label{eq:torsion}
\Gamma^\lambda_{[\mu\rho]}=-v^\lambda\partial_{[\mu}\tau_{\rho]}-h^{\nu\lambda}\tau_{[\rho}K_{\mu]\nu}-v^\lambda X_{[\mu\rho]}+\frac{1}{2}h^{\nu\lambda}Y_{\nu[\mu\rho]}\,.
\end{equation}
An alternative way of writing \eqref{eq:finalGamma} that makes manifest the property $\nabla_\mu v^\nu=0$ is as follows
\begin{eqnarray}
\Gamma^\lambda_{\mu\rho} & = & \tau_\rho\partial_\mu v^\lambda-h_{\rho\sigma}\partial_\mu h^{\sigma\lambda}+\frac{1}{2}h_{\rho\sigma}h^{\kappa\sigma}h^{\nu\lambda}\left(\partial_\kappa h_{\mu\nu}-\partial_\mu h_{\nu\kappa}-\partial_\nu h_{\mu\kappa}\right)\nonumber\\
&&-v^\lambda X_{\mu\rho}+\frac{1}{2}h^{\nu\lambda}Y_{\nu\mu\rho}\,.
\end{eqnarray}

The requirement is that $\Gamma^\rho_{\mu\nu}$ transforms as an affine connection under general coordinate transformations and remains inert under $C$ and $J$ (tangent space) transformations. The first line of \eqref{eq:finalGamma} transforms affinely, i.e. it has a term $\partial_\mu\partial_\rho\xi^\lambda$ plus terms that transform tensorially. In fact the last term of the first line containing the extrinsic curvature transforms as a tensor and is thus not responsible for producing the $\partial_\mu\partial_\rho\xi^\lambda$ term. This means that all terms on the second line of \eqref{eq:finalGamma} must transform as tensors, i.e. $X_{\mu\rho}$ and $Y_{\nu\mu\rho}$ transform as tensors under general coordinate transformations. 

As a check that we have in fact managed to write all the components of $\Omega_\mu{}^a$ and $\Omega_\mu{}^{ab}$ in terms of a Carrollian metric compatible $\Gamma^\rho_{\mu\nu}$ we count the number of components in $h_{\sigma\rho}X^\sigma_{[\nu\mu]}$ (since this determines $Y_{\nu\mu\rho}$ via equation \eqref{eq:Y}) and $X_{\mu\nu}$. The tensor $h_{\sigma\rho}X^\sigma_{[\nu\mu]}$ has $\frac{1}{2}(d+1)^2d-\frac{1}{2}d(d+1)=\frac{1}{2}d^2(d+1)$ components and it satisfies $\frac{1}{2}d(d+1)$ constraints $v^\nu h_{\sigma\rho}X^\sigma_{[\nu\mu]}+v^\nu h_{\sigma\mu}X^\sigma_{[\nu\rho]}=0$ giving $\frac{1}{2}(d+1)d^2-\frac{1}{2}d(d+1)=\frac{1}{2}(d-1)d(d+1)$ free components. The tensor $X_{\mu\nu}$ has $(d+1)^2$ components satisfying $(d+1)$ constraints $v^\nu X_{\mu\nu}$ giving $d(d+1)$ free components. Together this gives $d(d+1)((d-1)+2)/2=d(d+1)^2/2$ components which is also the number of components in $\Omega_\mu{}^a$ ($d(d+1)$ components) and $\Omega_\mu{}^{ab}$ ($(d+1)d(d-1)/2$ components). Equation \eqref{eq:XandboostOmega} expresses the relation between $X_{\mu\rho}$ and $\Omega_\mu{}^a$. Using that the vielbein postulates \eqref{eq:VP1} and \eqref{eq:VP2} imply
\begin{equation}
\Gamma^\rho_{\mu\nu} = -v^\rho\partial_\mu\tau_\nu+v^\rho\Omega_{\mu a}e^a_\nu+e^\rho_a\partial_\mu e_\nu^a-\Omega_\mu{}^a{}_b e^\rho_a e^b_\nu\,,
\end{equation}
we obtain the following relation between $\Omega_\mu{}^a{}_b$ and $h^{\rho\sigma}Y_{\sigma\mu\nu}$
\begin{equation}
\frac{1}{2}h^{\rho\sigma}Y_{\sigma\mu\nu} = -\frac{1}{2}h^{\rho\sigma}\left(\partial_\mu h_{\sigma\nu}+\partial_\nu h_{\mu\sigma}-\partial_\sigma h_{\mu\nu}\right)+h^{\rho\sigma}\tau_\nu K_{\mu\sigma}+e^\rho_a\partial_\mu e_\nu^a-\Omega_\mu{}^a{}_b e^\rho_a e^b_\nu\,.
\end{equation}

Finally in order that our $\Gamma^\rho_{\mu\nu}$ satisfies all the required properties we must ensure that it is invariant under local $C$ and $J$ transformations. It is manifestly $J$ invariant so we are left to ensure local $C$ invariance. Using that 
\begin{eqnarray}
\delta_C\tau_\mu & = & \lambda_\mu\,,\\
\delta_C h^{\mu\nu} & = & \left(h^{\mu\sigma}v^\nu+h^{\nu\sigma}v^\mu\right)\lambda_\sigma\,,\label{eq:Gtrafoinvh}
\end{eqnarray}
where $\lambda_\mu=e_\mu^a\lambda_a$, one can shown that $\Gamma^\rho_{\mu\nu}$ is $C$ invariant if and only if $X_{\mu\rho}$ and $Y_{\nu\mu\rho}$ transform as
\begin{eqnarray}
\delta_C X_{\mu\rho} & = & -\left(\partial_\mu\lambda_\rho-\Gamma^\sigma_{\mu\rho}\lambda_\sigma\right)\,,\label{eq:trafoX}\\
\delta_C Y_{\nu\mu\rho} & = & 2\lambda_\rho K_{\mu\nu}-2\lambda_\nu K_{\mu\rho}\,.\label{eq:trafoY}
\end{eqnarray}
These transformation rules are compatible with the properties $v^\rho X_{\mu\rho}=0$ and $v^\nu Y_{\nu\mu\rho}=v^\rho Y_{\nu\mu\rho}=0$. For the transformation of $X_{\mu\rho}$ this is by virtue of $\lambda_\nu\nabla_\mu v^\nu=0$ (i.e. metric compatibility). The transformation of $X_{\mu\rho}$ involves the connection $\Gamma^\rho_{\mu\nu}$. However it does not involve the tensor $X_{\mu\rho}$ on the right hand side of \eqref{eq:trafoX} because $\Gamma^\rho_{\mu\nu}$ is contracted with $\lambda_\rho$ which has the property that $v^\rho\lambda_\rho=0$. In fact we can rewrite the right hand side of \eqref{eq:trafoX} as follows. Using \eqref{eq:finalGamma} we find
\begin{equation}\label{eq:result1}
\Gamma^\sigma_{\mu\rho}\lambda_\sigma=\frac{1}{2}\left(\partial_\mu\lambda_\rho+\partial_\rho\lambda_\mu\right)-\frac{1}{2}\mathcal{L}_u h_{\mu\rho}+\frac{1}{2}u^\nu\tau_\rho \mathcal{L}_vh_{\mu\nu}+\frac{1}{2}u^\nu Y_{\nu\mu\rho}\,,
\end{equation}
where we defined the vector $u^\mu=h^{\mu\sigma}\lambda_\sigma$ and where we used \eqref{eq:extrinsiccurv}. Using that
\begin{equation}
u^\nu \mathcal{L}_vh_{\mu\nu}=v^\nu\left(\partial_\nu\lambda_\mu-\partial_\mu\lambda_\nu\right)-v^\nu\mathcal{L}_u h_{\mu\nu}\,,
\end{equation}
we obtain for $\delta_C X_{\mu\rho}$ the result
\begin{equation}\label{eq:finaltrafoX}
\delta_C X_{\mu\rho}=-\frac{1}{2}\left(\delta^\nu_\rho+\tau_\rho v^\nu\right)\left(\partial_\mu\lambda_\nu-\partial_\nu\lambda_\mu+\mathcal{L}_u h_{\mu\nu}\right)+\frac{1}{2}u^\kappa Y_{\kappa\mu\rho}\,,
\end{equation}
making it manifest that $\delta_C \left(v^\rho X_{\mu\rho}\right)=v^\rho\delta_C X_{\mu\rho}=0$. One may wonder why there is a term transforming into $u^\kappa Y_{\kappa\mu\rho}$. The reason is that the transformation of $h^{\nu\lambda}Y_{\nu\mu\rho}$ in \eqref{eq:finalGamma} produces such terms through \eqref{eq:Gtrafoinvh} and these need to be cancelled. 

Using \eqref{eq:result1} and \eqref{eq:extrinsiccurv} we can also write the variation of $X_{\mu\rho}$ in the following manner
\begin{equation}\label{eq:deltaGX}
\delta_C X_{\mu\rho}=-\frac{1}{2}\left(\partial_\mu\lambda_\rho-\partial_\rho\lambda_\mu\right)-\frac{1}{2}\mathcal{L}_u h_{\mu\rho}+\frac{1}{2}u^\nu \left(Y_{\nu\mu\rho}-2\tau_\rho K_{\mu\nu}+2\tau_\nu K_{\mu\rho}\right)\,.
\end{equation}
This way of writing $\delta_C X_{\mu\rho}$ is useful when one tries to write the right hand side as the $\delta_C$ of something which we will do in the next subsection. The term $u^\nu\tau_\nu K_{\mu\rho}$ has been added to make manifest that $u^\nu$ contracts a term that is $C$-boost invariant. Of course because $u^\nu\tau_\nu=0$ the added term vanishes. If we write in \eqref{eq:deltaGX} and likewise in \eqref{eq:trafoY} the parameter $\lambda_\mu=h_{\mu\nu}u^\nu$ then $u^\mu$ always contracts or multiplies a term that is manifestly Carrollian boost invariant. This is not the case for the parameter $\lambda_\mu$ because it sometimes is contracted with $h^{\mu\nu}$ which is not invariant under local $C$ transformations.

\subsection{Introducing the vector $M^\mu$}\label{subsec:vectorM}

So far we have considered the most general case where the $\bar\delta$ transformations are realized on the set of fields $\tau_\mu$, $e^a_\mu$, $\Omega_\mu{}^a$ and $\Omega_\mu{}^{ab}$ or what is the same $\tau_\mu$, $e^a_\mu$ and $\Gamma^\rho_{\mu\nu}$ where the latter is metric compatible in the sense that $\nabla_\mu v^\nu=\nabla_\mu h_{\nu\rho}=0$. In the remainder we will realize the algebra on a smaller set of fields. 

Sometimes when gauging algebras, as happens e.g. when gauging the Poincar\'e algebra, it is possible to realize the $\bar\delta$ transformations on a smaller set of fields by imposing curvature constraints whose effect is to make some of the connections in $\mathcal{A}_\mu$ dependent on other connections in $\mathcal{A}_\mu$. For example in the case of the gauging of the Poincar\'e algebra setting the torsion to zero, i.e. imposing the curvature constraint $R_{\mu\nu}{}^a(P)=0$ (where $P$ denotes the space-time translations), enables one to express the spin connection coefficients $\omega_\mu{}^{ab}$ in terms of $e_\mu^a$. 

In the case of the gauging of the Bargmann algebra imposing curvature constraints (without introducing new fields) to write the Galilean boost and spatial rotation connections in terms of the vielbeins and the central charge gauge connection is only possible when there is no torsion \cite{Andringa:2010it}. When there is torsion the curvature constraints become dependent on an additional St\"uckelberg scalar field $\chi$ that is not present in $\mathcal{A}_\mu$. This field needs to be added to ensure the correct transformation properties of the Galilean boost and spatial rotation connections when writing them as dependent gauge connections \cite{Hartong:2014oma,Bergshoeff:2014uea,Hartong:2015zia}. 
In the context of formulating Ho\v rava--Lifshitz (HL) gravity as a theory of dynamical torsional Newton--Cartan geometry \cite{Hartong:2015zia} the St\"uckelberg scalar field $\chi$ plays an important role in making the identification between TNC and HL variables. In the context to Ho\v rava--Lifshitz gravity this field was introduced in \cite{Horava:2010zj} and dubbed the Newtonian prepotential.

In the case of field theory on Newton--Cartan space-times including torsion is crucial because it allows one to compute the energy current \cite{Christensen:2013lma,Christensen:2013rfa,Geracie:2014nka,Hartong:2014oma,Hartong:2014pma}. The fact that one needs to introduce an extra St\"uckelberg scalar field to the formalism when there is torsion does not mean that any field theory on such a background has a non-trivial response to varying the St\"uckelberg scalar. It can happen that there are additional local symmetries in the model that allow one to remove this field from the action \cite{Hartong:2014pma,Hartong:2015wxa}. The main message is that once we start imposing curvature constraints the resulting reduced set of fields on which the algebra is realized do not need to correspond to a constrained algebra gauging and may involve new fields.

In both the gauging of the Poincar\'e algebra and of the Bargmann algebra the effect of the curvature constraints is to make the connection $\Gamma^\rho_{\mu\nu}$ a fully dependent connection. Imposing the curvature constraint $R_{\mu\nu}{}^a(P)=0$ in the Poincar\'e case leads to the Levi-Civit\`a connection\footnote{There also exists the possibility to set the Riemann curvature 2-form $R_{\mu\nu}{}^{ab}(M)$, where $M$ is the generator of Lorentz transformations, equal to zero. This leads to the so-called Weitzenb\"ock connection (see for example \cite{Ortin:2004ms}). We refer to \cite{Hofman:2014loa} for similar ideas in the context of gauging the Carroll algebra.}

In the case of the gauging of the Bargmann algebra in the presence of torsion the algebra of $\bar\delta$ transformations is realized on $\tau_\mu$, $e_\mu^a$ and $M_\mu=m_\mu-\partial_\mu\chi$. One can also say that from the point of view of gauging the Galilei algebra one needs to add the vector $M_\mu$ to construct a $\Gamma^\rho_{\mu\nu}$ that obeys all the properties of an affine connection \cite{Hartong:2015zia}. In other words from the point of view of adding curvature constraints to the gauging of the Galilei algebra we add a vector $M_\mu$ with appropriately chosen transformation properties to realize the algebra on the smaller number of fields $\tau_\mu$, $e_\mu^a$ and $M_\mu$ as opposed to $\tau_\mu$, $e_\mu^a$, $\Omega_\mu{}^a$ (local Galilean boosts) and $\Omega_\mu{}^{ab}$ (local spatial rotations). 

In the case discussed here we will realize the algebra of $\bar\delta$ transformations on $\tau_\mu$, $e^a_\mu$ and a contravariant vector field $M^\mu$ where $M^\mu$ transforms under the $\bar\delta$ transformations as 
\begin{equation}\label{eq:bardeltaM}
\bar\delta M^\mu=\mathcal{L}_\xi M^\mu+e^\mu_a\lambda^a=\mathcal{L}_\xi M^\mu+h^{\mu\nu}\lambda_\mu=\mathcal{L}_\xi M^\mu+u^\mu\,.
\end{equation}
We are not aware of an extension of the Carroll algebra such that $M^\mu$ can be constructed from the additional connections appearing in $\mathcal{A}_\mu$ corresponding to the extended Carroll algebra. The guiding principle will be to write $\Gamma^\rho_{\mu\nu}$ in terms of $\tau_\mu$, $e^a_\mu$ and $M^\mu$ in such a way that it obeys all the required properties. In other words we need to write the tensors $X_{\mu\rho}$ and $Y_{\sigma\mu\nu}$ in terms of $\tau_\mu$, $e^a_\mu$ and $M^\mu$ ensuring that they transform correctly under the $\bar\delta$ transformations. A raison d'\^etre for the vector $M^\mu$ will be given in the next section.

One of the benefits of working with $X_{\mu\rho}$ and $Y_{\sigma\mu\nu}$ is that their transformation properties under local tangent space $C$ and $J$ transformations is much simpler than for the equivalent set of objects $\Omega_\mu{}^a$ and $\Omega_\mu{}^{ab}$. Both $X_{\mu\rho}$ and $Y_{\sigma\mu\nu}$ are invariant under $J$ transformations and their the $C$ transformations are given in \eqref{eq:trafoX} (or equivalently \eqref{eq:deltaGX}) and \eqref{eq:trafoY}. We will now use the additional $M^\mu$ vector to write down a realization of $X_{\mu\rho}$ and $Y_{\sigma\mu\nu}$ in terms of $\tau_\mu$, $e^a_\mu$ and $M^\mu$. 

Using \eqref{eq:deltaGX} and \eqref{eq:bardeltaM} we can write
\begin{eqnarray}
0 & = & \delta_C\left(X_{\mu\rho}+\frac{1}{2}\partial_\mu \left(h_{\rho\sigma}M^\sigma\right)-\frac{1}{2}\partial_\rho\left(h_{\mu\sigma}M^\sigma\right)+\frac{1}{2}\mathcal{L}_M h_{\mu\rho}\right.\nonumber\\
&&\left.-\frac{1}{2}M^\nu \left(Y_{\nu\mu\rho}-2\tau_\rho K_{\mu\nu}+2\tau_\nu K_{\mu\rho}\right)\right)\,.
\end{eqnarray}
Hence a realization of $X_{\mu\rho}$ (but not the most general one) is to write
\begin{eqnarray}
X_{\mu\rho} & = & -\frac{1}{2}\partial_\mu \left(h_{\rho\sigma}M^\sigma\right)+\frac{1}{2}\partial_\rho\left(h_{\mu\sigma}M^\sigma\right)-\frac{1}{2}\mathcal{L}_M h_{\mu\rho}\nonumber\\
&&+\frac{1}{2}M^\nu \left(Y_{\nu\mu\rho}-2\tau_\rho K_{\mu\nu}+2\tau_\nu K_{\mu\rho}\right)\,,\label{eq:X2}
\end{eqnarray}
obeying $v^\rho X_{\mu\rho}=0$. Likewise for $Y_{\nu\mu\rho}$ we can take
\begin{equation}\label{eq:Y2}
Y_{\nu\mu\rho} = 2h_{\rho\sigma}M^\sigma K_{\mu\nu}-2h_{\nu\sigma}M^\sigma K_{\mu\rho}\,,
\end{equation}
which transforms as in \eqref{eq:trafoY} and obeys $v^\nu Y_{\nu\mu\rho}=v^\rho Y_{\nu\mu\rho}=0$. Substituting \eqref{eq:X2} and \eqref{eq:Y2} into \eqref{eq:finalGamma} we obtain
\begin{equation}\label{eq:manifestinvGamma}
\Gamma^\lambda_{\mu\rho}=-v^\lambda\partial_\mu\hat\tau_\rho+\frac{1}{2}\bar h^{\nu\lambda}\left(\partial_\mu h_{\rho\nu}+\partial_\rho h_{\mu\nu}-\partial_\nu h_{\mu\rho}\right)-\bar h^{\nu\lambda}\hat\tau_\rho K_{\mu\nu}+\bar h^{\nu\lambda}\hat\tau_\nu K_{\mu\rho}\,,
\end{equation}
where we defined 
\begin{eqnarray}
\hat\tau_\mu & = & \tau_\mu-h_{\mu\nu}M^\nu\,,\label{eq:hattau}\\
\bar h^{\mu\nu} & = & h^{\mu\nu}-M^\mu v^\nu-M^\nu v^\mu\,,\label{eq:barinvh}
\end{eqnarray}
which are manifestly $C$ invariant. Another $C$ invariant (scalar) quantity that we can define is $\bar\Phi$ which is given by
\begin{equation}\label{eq:barPhi}
\bar\Phi=-M^\nu\tau_\nu +\frac{1}{2}h_{\nu\sigma}M^\nu M^\sigma\,.
\end{equation}
The affine connection \eqref{eq:manifestinvGamma} has the property that if we replace all $\bar h^{\mu\nu}$ by $H^{\mu\nu}=\bar h^{\mu\nu}+\alpha\bar\Phi v^\mu v^\nu$ the resulting expression for $\Gamma^\lambda_{\mu\rho}$ remains unchanged, i.e. does not depend on $\alpha$. Hence we can take $\alpha=2$ and write for $\Gamma^\lambda_{\mu\rho}$ in \eqref{eq:manifestinvGamma}
\begin{equation}\label{eq:manifestinvGamma2}
\Gamma^\lambda_{\mu\rho}=-v^\lambda\partial_\mu\hat\tau_\rho+\frac{1}{2}\hat h^{\nu\lambda}\left(\partial_\mu h_{\rho\nu}+\partial_\rho h_{\mu\nu}-\partial_\nu h_{\mu\rho}\right)-\hat h^{\nu\lambda}\hat\tau_\rho K_{\mu\nu}\,,
\end{equation}
where $\hat h^{\nu\lambda}$ is defined by
\begin{equation}
\hat h^{\nu\lambda}=\bar h^{\nu\lambda}+2\bar\Phi v^\nu v^\lambda\,,
\end{equation}
for which $\hat\tau_\mu\hat h^{\mu\nu}=0$. The connection \eqref{eq:manifestinvGamma2} is independent of $\bar\Phi$ because it can be shown that $M^\mu$ appears in $\hat\tau_\mu$ and $\hat h^{\mu\nu}$ only via $h_{\mu\nu}M^\nu$. This is made more manifest below following the discussion around equations \eqref{eq:hate} and \eqref{eq:invg}.

The connection \eqref{eq:manifestinvGamma} satisfies by design the metric compatibility conditions $\nabla_\mu v^\nu=\nabla_\mu h_{\nu\rho}=0$. However it also satisfies the conditions
\begin{equation}\label{eq:extraprops}
\nabla_\mu\hat\tau_\nu=\nabla_{\mu}\hat h^{\nu\rho}=0\,,
\end{equation}
where $\nabla_\mu\hat\tau_\nu$ follows immediately by inspection of \eqref{eq:manifestinvGamma} using that $\hat h^{\mu\nu}\hat\tau_\nu=0$. The second property $\nabla_{\mu}\hat h^{\nu\rho}=0$ follows from all the other metric compatibility conditions and the fact that $\hat h^{\nu\rho}$ is fully determined once $\hat \tau_\mu$ and $h_{\mu\nu}$ are known. The property $\nabla_\mu\hat\tau_\nu=0$ implies that $\nabla_\mu\tau_\rho=\nabla_\mu\left(h_{\rho\sigma}M^\sigma\right)=-X_{\mu\rho}$ where we used \eqref{eq:hattau} and \eqref{eq:XandboostOmega} and is compatible with the transformation under local $C$ transformations given in \eqref{eq:trafoX}. We stress though that the properties \eqref{eq:extraprops} are special for the particular realization of $\Gamma^\lambda_{\mu\rho}$ given in \eqref{eq:manifestinvGamma} and will not be true for other realizations of $\Gamma^\lambda_{\mu\rho}$ that for example also depend on the scalar invariant $\bar\Phi$.

We can define a new set of vielbeins $\hat\tau_\mu$, $e_\mu^a$ whose inverse is $v^\mu$, $\hat e^\mu_a$ with the latter defined by
\begin{equation}\label{eq:hate}
\hat e^\mu_a=e^\mu_a-M^\rho e_{\rho a} v^\mu\,.
\end{equation}
These new vielbeins satisfy
\begin{equation}
v^\mu\hat\tau_\mu=-1\,,\qquad v^\mu e^a_\mu=0\,,\qquad \hat e^\mu_a\hat\tau_\mu=0\,,\qquad \hat e^\mu_ae^b_\mu=\delta^b_a\,.
\end{equation}
Out of these objects we can build a Lorentzian symmetric rank two tensor $g_{\mu\nu}$ via
\begin{equation}\label{eq:g}
g_{\mu\nu}=-\hat\tau_\mu\hat\tau_\nu+h_{\mu\nu}=-\hat\tau_\mu\hat\tau_\nu+\delta_{ab}e^a_\mu e^b_\nu\,,
\end{equation}
whose inverse is
\begin{equation}\label{eq:invg}
g^{\mu\nu}=-v^\mu v^\nu+\hat h^{\mu\nu}=-v^\mu v^\nu+\delta^{ab}\hat e^\mu_a \hat e^\nu_b\,.
\end{equation}

Since the connection \eqref{eq:manifestinvGamma} satisfies $\nabla_\mu\hat\tau_\nu=\nabla_\mu h_{\nu\rho}=0$ it in particular obeys $\nabla_\mu g_{\nu\rho}=0$. Since it furthermore has torsion the connection \eqref{eq:manifestinvGamma} must be a special case of a Riemann--Cartan connection. By this we mean a torsionful connection obeying $\nabla_\mu g_{\nu\rho}=0$. Any such connection must be of the form \cite{Ortin:2004ms}
\begin{equation}
\Gamma^\lambda_{\mu\rho}=\frac{1}{2}g^{\nu\lambda}\left(\partial_\mu g_{\rho\nu}+\partial_\rho g_{\mu\nu}-\partial_\nu g_{\mu\rho}\right)+g^{\nu\lambda}\left(\Gamma^\kappa_{[\nu\rho]}g_{\kappa\mu}+\Gamma^\kappa_{[\nu\mu]}g_{\kappa\rho}+\Gamma^\kappa_{[\mu\rho]}g_{\kappa\nu}\right)\,.
\end{equation}
By using \eqref{eq:g} and \eqref{eq:invg} we can show that \eqref{eq:manifestinvGamma} is of this form with torsion given by
\begin{equation}
\Gamma^\lambda_{[\mu\rho]}=-v^\lambda\partial_{[\mu}\hat\tau_{\rho]}-\hat h^{\nu\lambda}\hat\tau_{[\rho}K_{\mu]\nu}\,,
\end{equation}
compatible with \eqref{eq:torsion}.

The connection \eqref{eq:manifestinvGamma} is not the most general affine connection compatible with our requirements. We still have the freedom to add to $X_{\mu\rho}$ and $Y_{\nu\mu\rho}$ terms that are invariant under local Carrollian boosts. When we add a term to $Y_{\nu\mu\rho}$ we should also add the corresponding term to \eqref{eq:X2} because $\frac{1}{2}M^\nu Y_{\nu\mu\rho}$ appears in $X_{\mu\rho}$. Further any term added to $Y_{\nu\mu\rho}$ must obey the property that when contracted with $v^\nu$ or $v^\rho$ it vanishes since $Y_{\nu\mu\rho}$ obeys this property. Equation \eqref{eq:manifestinvGamma2} is independent of $\bar\Phi$ and the only terms that we can still add to $\Gamma^\rho_{\mu\nu}$ without affecting its properties come from $\bar\Phi$ dependent terms that we add to $X_{\mu\rho}$ and $Y_{\nu\mu\rho}$. An example of such a term is to add to $X_{\mu\rho}$ a term proportional to $\bar\Phi K_{\mu\rho}$ which is $C$ invariant and orthogonal to $v^\rho$. The effect is to redefine $\Gamma^\lambda_{\mu\rho}$ by a term proportional to $\bar\Phi v^\lambda K_{\mu\rho}$. Yet another term that we can add to $X_{\mu\rho}$ compatible with $\Gamma^\lambda_{\mu\rho}$ remaining invariant under $C$, $J$ transformations, transforming affinely, and being metric compatible in the Carrollian sense, is a term proportional to $h_{\mu\rho}v^\sigma\partial_\sigma\bar\Phi$. Any of these affine connections is an allowed connection and so one can choose them to suit one's convenience. The same phenomenon happens for the case of torsional Newton--Cartan (TNC) geometry. Sometimes it is useful to work with a TNC connection that does not depend on the scalar $\tilde\Phi$ (the TNC counterpart of $\bar\Phi$ defined in \eqref{eq:tildePhi}) as is for example the case when making contact with Ho\v rava--Lifshitz gravity \cite{Hartong:2015zia} and sometimes it is useful to work with a TNC connection depending linearly on $M_\mu$ as is for example the case when coupling field theories with particle number symmetry to TNC backgrounds \cite{Jensen:2014aia,Hartong:2014pma,Hartong:2015wxa}.

\section{The Geometry on Null Hypersurfaces}\label{sec:nullhypersurfaces}

A natural example of a space-time with a Carrollian metric structure is a null hypersurface embedded into a Lorentzian space-time of one dimension higher \cite{Duval:2014uoa,Duval:2014uva}. Before introducing a Carrollian space-time as the geometry on a null hypersurface it is useful to consider first the case of a Newton--Cartan space-time as the geometry orthogonal to a null Killing vector. This will also enable us to compare the two cases later.

\subsection{Newton--Cartan space-time}\label{subsec:NCspace}

It is well known that Newton--Cartan geometry on a manifold with coordinates $x^\mu$ can be obtained by null reduction \cite{Duval:1984cj,Julia:1994bs,Bekaert:2013fta,Christensen:2013lma,Christensen:2013rfa}, i.e. by starting from a Lorentzian space-time with one extra dimension $u$ whose metric is of the form
\begin{eqnarray}
ds^2 & = & 2\tau_\mu dx^\mu\left(du-m_\nu dx^\nu\right)+h_{\mu\nu}dx^\mu dx^\nu\nonumber\\
& = & 2\tau_\mu dx^\mu du+\bar h_{\mu\nu}dx^\mu dx^\nu\,,\label{eq:TNCembedding}
\end{eqnarray}
where we take $\partial_u$ to be a Killing vector so that $\tau_\mu$ and $\bar h_{\mu\nu}$ are independent of $u$ and where 
\begin{equation}\label{eq:barh}
\bar h_{\mu\nu}=h_{\mu\nu}-m_\mu \tau_\nu-m_\nu \tau_\mu\,.
\end{equation}
Note that for this metric we have $g_{uu}=0$. The inverse metric components are
\begin{equation}
g^{\mu\nu}=h^{\mu\nu}\,,\qquad g^{\mu u}=-\hat v^\mu\,,\qquad g^{uu}=2\tilde\Phi\,,
\end{equation}
where
\begin{eqnarray}
\hat v^\mu & = & v^\mu-h^{\mu\nu}m_\nu\,,\label{eq:hatv}\\
\tilde\Phi & = & -v^\mu m_\mu+\frac{1}{2}h^{\mu\nu}m_\mu m_\nu\,.\label{eq:tildePhi}
\end{eqnarray}
The metric \eqref{eq:TNCembedding} is the most general Lorentzian metric with a null Killing vector $\partial_u$. The coordinate transformations that preserve the form of the null Killing vector are
\begin{eqnarray}
u' & = & u-\sigma(x)\,,\\
x'^\mu & = & x'^\mu(x)\,.
\end{eqnarray}
Under the shift in $u$ the vector $m_\mu$ transforms as a $U(1)$ connection
\begin{equation}
m'_\mu=m_\mu-\partial_\mu\sigma\,.
\end{equation}
A TNC metric compatible connection can be found by taking the Levi-Civit\`a connection of the higher dimensional space-time with all its legs in the $x^\mu$ directions and to add torsion to this by hand so as to make it metric compatible in the TNC sense, i.e. $\nabla_\mu\tau_\nu=\nabla_\mu h^{\nu\rho}=0$ \cite{Christensen:2013lma,Christensen:2013rfa}. Instead of speaking about null reduction, one can say that TNC geometry is the geometry on the space-time orthogonal to the null Killing vector $\partial_u$.

If we insist that the connection on the TNC space-time is naturally induced from the Levi-Civit\`a connection on the higher dimensional space-time we need to impose that $\partial_u$ is hypersurface orthogonal. To see this write for the higher dimensional metric 
\begin{equation}\label{eq:nullbeins1}
ds^2=g_{AB}dx^A dx^B=\left(U_A V_B+U_B V_A+\Pi_{AB}\right)dx^Adx^B\,,
\end{equation}
where $x^A=(u,x^\mu)$. The vectors $U_A$ and $V_A$ are nullbeins defined by
\begin{equation}\label{eq:nullbeins2}
U^AU_A=0\,,\qquad V^AV_A=0\,,\qquad U^A V_A=-1\,,\qquad U^A \Pi_{AB}=V^A\Pi_{AB}=0\,.
\end{equation}
We choose $U^A=(\partial_u)^A$, so that
\begin{equation}
U_u=0\,,\quad U_\mu=-\tau_\mu\,,\quad V_u=-1\,,\quad V_\mu=m_\mu\,,\quad \Pi_{uA}=0\,,\quad\Pi_{\mu\nu}=h_{\mu\nu}\,,
\end{equation}
and 
\begin{equation}
V^u=-v^\mu m_\mu\,,\quad V^\mu=-v^\mu\,,\quad \Pi^{uu}=h^{\mu\nu}m_\mu m_\nu\,,\quad \Pi^{u\mu}=h^{\mu\nu}m_\nu\,,\quad \Pi^{\mu\nu}=h^{\mu\nu}\,.
\end{equation}
In order that we have a TNC connection on the space-time orthogonal to $U^A$ we demand that $\nabla_A U_B$ projected along all directions orthogonal to $U^A$ gives zero, i.e.
\begin{eqnarray}
U^A U^B\nabla_A U_B & = & 0\,,\label{eq:1}\\
U^A\Pi^B{}_C\nabla_A U_B & = & 0\,,\label{eq:2}\\
\Pi^A{}_C U^B\nabla_A U_B & = & 0\,,\label{eq:3}\\
\Pi^A{}_C\Pi^B{}_D\nabla_A U_B & = & 0\,,\label{eq:4}
\end{eqnarray}
where $\nabla_A$ contains the Levi-Civit\`a connection. These conditions lead to the TNC metric compatibility condition $\nabla_\mu\tau_\nu=0$. Likewise to obtain $\nabla_\mu h^{\nu\rho}=0$ we impose that $\nabla_A \Pi^{BC}$ with all indices projected onto directions orthogonal to $U^A$ gives zero. Since we have 
\begin{equation}
\nabla_A \Pi^{BC}=-U^B\nabla_A V^C-V^C\nabla_A U^B-U^C\nabla_A V^B-V^B\nabla_A U^C\,,
\end{equation}
which follows from the fact that $\nabla_A g^{BC}=0$ we obtain
\begin{eqnarray}
\Pi^D{}_B \Pi^E{}_C\nabla_A \Pi^{BC} & = & 0\,,\\
U_B U_C\nabla_A \Pi^{BC} & = & 0\,,\\
\Pi^D{}_B U_C\nabla_A \Pi^{BC} & = & -\Pi^{DC}\nabla_A U_C\,.
\end{eqnarray}
Hence to enforce $\nabla_\mu h^{\nu\rho}=0$ we only need that $\Pi^{DC}\nabla_A U_C$ contracted with $U^A$ and $\Pi^A{}_B$ gives zero. These conditions are already imposed 
in \eqref{eq:2} and \eqref{eq:4}. Since $U^A$ is a null Killing vector equations \eqref{eq:1} to \eqref{eq:3} together with the symmetric part of \eqref{eq:4} are satisfied. What remains is to impose that the spatial projection of the antisymmetric part of $\nabla_A U_B$ vanishes which is equivalent to demanding that $U_A$ is hypersurface orthogonal. Put another way it must be that 
\begin{equation}\label{eq:inducedTNC}
\nabla_A U_B=U_A X_B-U_BX_A\,,
\end{equation}
for some vector $X_A$ obeying $U^AX_A=0$ (as follows from the fact that $U^A$ is a null Killing vector and thus geodesic) but otherwise arbitrary in order that the Levi-Civit\`a connection induces a Newton--Cartan connection on the space-time orthogonal to $\partial_u$. Since the left hand side of \eqref{eq:inducedTNC} is just $\tfrac{1}{2}\left(\partial_AU_B-\partial_B U_A\right)$ and $X_u=0$ the only non-trivial component of \eqref{eq:inducedTNC} is when $A=\mu$ and $B=\nu$ expressing the fact that $\tau_\mu$ is hypersurface orthogonal, but not necessarily closed. This is the case referred to as twistless torsional Newton--Cartan geometry (TTNC) \cite{Christensen:2013lma,Christensen:2013rfa}. In this case metric compatibility $\nabla_\mu\tau_\nu=0$ requires a torsionful connection. We can thus obtain a torsionful connection from a Riemannian geometry by projecting along directions orthogonal to a hypersurface orthogonal null Killing vector. 

One may wonder how this is possible since the connection of the Riemannian space-time is symmetric. From the properties of $U_A$ we infer that
\begin{eqnarray}
\partial_\mu\tau_\nu+\partial_\nu\tau_\mu & = & 2\Gamma^{\rho}_{(g)\mu\nu}\tau_\rho\,,\\
\partial_\mu\tau_\nu-\partial_\nu\tau_\mu & = & 2\tau_\mu X_\nu-2\tau_\nu X_\mu\,,
\end{eqnarray}
where $\Gamma^{\rho}_{(g)\mu\nu}$ is the Levi-Civit\`a connection with all indices in the $x^\mu$ directions. From the first of these two equations we read off that the symmetric part of TNC connection satisfies $\Gamma^\rho_{(\mu\nu)}\tau_\rho=\Gamma^{\rho}_{(g)\mu\nu}\tau_\rho$. In order to repackage these equations into $\nabla_\mu\tau_\nu=0$ we see that $X_\mu$ contributes to a torsion tensor $\Gamma^{\rho}_{[\mu\nu]}\tau_\rho=\tau_\mu X_\nu-\tau_\nu X_\mu$. In other words the torsion can be introduced due to the fact that we are dealing with a geometry orthogonal to a null vector $U^A$ so that there is a certain arbitrariness encoded in $X_A$ when solving for \eqref{eq:4}. The torsion is thus described by a vector $X_\mu$. In \cite{Hartong:2015zia} the torsion vector is denoted by $a_\mu$ which relates to $X_\mu$ via $a_\mu=-2X_\mu$. It determines whether we are dealing with projectable or non-projectable Ho\v rava--Lifshitz gravity.

The conditions \eqref{eq:inducedTNC} together with $U_A$ being a null Killing vector guarantee that a TTNC metric compatible $\Gamma^\rho_{\mu\nu}$ exists but the projection equations onto the space-time orthogonal to $U^A$ do not tell one the precise form of this connection. This is to be expected since there is a certain arbitrariness in the expression for $\Gamma^\rho_{\mu\nu}$.

We recall that in order to write an expression for $\Gamma^\rho_{\mu\nu}$ in terms of the TNC fields $\tau_\mu$, $e_\mu^a$ and $m_\mu$ that does not refer to an embedding in a higher dimensional space-time we need to add a St\"uckelberg scalar $\chi$ when there is torsion \cite{Hartong:2014oma,Bergshoeff:2014uea,Hartong:2015zia}. This amounts to replacing everywhere $m_\mu$ by $M_\mu=m_\mu-\partial_\mu\chi$.

\subsection{Carrollian space-time}\label{subsec:Cspace}

To obtain an embedding of a Carrollian space-time into a Lorentzian space-time of one dimension higher all that is required is to do the same as for the Newton--Cartan case but with the difference that it is now the inverse metric for which we take $g^{uu}=0$. In other words we write down the most general metric for which $g^{uu}=0$. Such a metric is given by
\begin{equation}\label{eq:embedCarr}
ds^2=du\left(2\bar\Phi du-2\hat\tau_\mu dx^\mu\right)+h_{\mu\nu}dx^\mu dx^\nu\,,
\end{equation}
where $\bar\Phi$ is given in \eqref{eq:barPhi} and $\hat\tau_\mu$ is given in \eqref{eq:hattau}.
The components of the inverse metric are
\begin{equation}
g^{uu}=0\,,\qquad g^{\mu u}=v^\mu\,,\qquad g^{\mu\nu}=\bar h^{\mu\nu}\,,
\end{equation}
where $\bar h^{\mu\nu}$ is given by \eqref{eq:barinvh}. The Carrollian space-time can be thought of as the geometry on the null hypersurface $u=\text{cst}$ whose normal is $\partial_A u$, i.e. it is the geometry orthogonal to $\partial_A u$. When $\tilde\Phi=\bar\Phi=0$ the Newton--Cartan and Carrollian geometry are the same. This is because the metrics \eqref{eq:TNCembedding} and \eqref{eq:embedCarr} become identical. One simply has the correspondence 
\begin{equation}\label{eq:swappingTNCandC}
\tau_\mu\leftrightarrow\hat\tau_\mu\,,\qquad\hat v^\mu\leftrightarrow v^\mu\,,\qquad\hat h_{\mu\nu}\leftrightarrow h_{\mu\nu}\,,\qquad h^{\mu\nu}\leftrightarrow\hat h^{\mu\nu}\,.
\end{equation}
In section \ref{subsec:comparing} we will discuss in more detail the relation between TNC and Carrollian geometry.

The coordinate transformations that preserve the null foliation are given by
\begin{eqnarray}
 u & = & u'\,,\\
 x^\mu & = & x^\mu(u',x')\,.
\end{eqnarray}
Under this transformation the vector $M^\mu$ transforms as
\begin{equation}
M'^\mu=M^\nu\frac{\partial x'^\mu}{\partial x^\nu}+\frac{\partial x'^\mu}{\partial u}\,.
\end{equation}
If we demand that $\partial_u$ is a Killing vector the coordinate transformations cannot depend on $u$ so that $M^\mu$ simply transforms as a vector. Alternatively if we work at a fixed value of $u$, i.e. a specific null hypersurface, the coordinate transformation of $x^\mu$ cannot depend on $u$ either and again $M^\mu$ transforms as a vector on the $u=\text{cst}$ hypersurface. We thus see from the embedding point of view that there is no extra symmetry associated with the vector $M^\mu$ while there is one in the NC case where we had a $U(1)$ acting on $m_\mu$ corresponding to the Bargmann extension of the Galilei algebra. 

We now discuss under what conditions the Carrollian metric compatible connection can be obtained from the Levi-Civit\`a connection in the higher dimensional space-time. To this end consider again \eqref{eq:nullbeins1} and \eqref{eq:nullbeins2}. This time we choose $U_A=\partial_A u$ implying that
\begin{equation}
U^u=0\,,\quad U^\mu=-v^\mu\,,\quad V^u=-1\,,\quad V^\mu=M^\mu\,,\quad \Pi^{uA}=0\,,\quad \Pi^{\mu\nu}=h^{\mu\nu}\,,
\end{equation}
as well as
\begin{equation}
\Pi_{u\mu}=h_{\mu\nu}M^\nu\,,\quad \Pi_{uu}=h_{\mu\nu}M^\mu M^\nu\,,\quad \Pi_{\mu\nu}=h_{\mu\nu}\,,\quad V_u=-\tau_\mu M^\mu\,,\quad V_\mu=-\tau_\mu\,.
\end{equation}
Imposing that $\nabla_\mu v^\nu=0$ amounts to demanding that
\begin{eqnarray}
U^A U_B\nabla_A U^B & = & 0\,,\\
\Pi^A{}_C U_B\nabla_A U^B & = & 0\,,\\
U^A \Pi^C{}_B\nabla_A U^B & = & 0\,,\\
\Pi^A{}_C \Pi^D{}_B\nabla_A U^B & = & 0\,.
\end{eqnarray}
The first and the second conditions are satisfied because $U^A$ is null while the third is satisfied because we furthermore know that $\nabla_A U_B=\nabla_B U_A$ due to our choice of $U_A$ as $\partial_A u$. The most general expression for $\nabla_A U_B$ compatible with all of the above conditions and the properties of $U_A$ is given by
\begin{equation}\label{eq:5}
\nabla_A U_B=U_A X_B+U_B X_A\,,
\end{equation}
where $X_A$ satisfies $U^A X_A=0$ but is otherwise an arbitrary vector. Using that $\nabla_A g_{BC}=0$, i.e. that
\begin{equation}
\nabla_A\Pi_{BC}=-U_B\nabla_A V_C-U_C\nabla_A V_B-V_B\nabla_A U_C-V_C\nabla_A U_B\,,
\end{equation}
we find that 
\begin{eqnarray}
\Pi^B{}_D\Pi^C{}_E\nabla_A\Pi_{BC} & = & 0\,,\\
U^B U^C\nabla_A\Pi_{BC} & = & 0\,,\\
\Pi^B{}_D U^C\nabla_A\Pi_{BC} & = & -U_A\Pi_D{}^C X_C\,,
\end{eqnarray}
where in the last relation we used \eqref{eq:5}. Hence $\nabla_A\Pi_{BC}$ vanishes when projected along directions orthogonal to $U_A$. Therefore, whereas in the NC case we had to demand equation \eqref{eq:inducedTNC} in the Carrollian case we need that \eqref{eq:5} holds in order that the induced connection comes from the Levi-Civit\`a connection of the higher dimensional space-time.

\subsection{Comparing Newton--Cartan and Carrollian space-times}\label{subsec:comparing}

As one can notice by comparing the discussions of sections \ref{subsec:NCspace} and \ref{subsec:Cspace} there are strong similarities between the geometry of TNC and Carrollian space-times. In fact in \cite{Duval:2014uoa} a certain duality between the two geometries has been proposed. Here we will extend this duality to include the TNC vector $M_\mu$ and the Carrollian vector $M^\mu$. The TNC metric-like objects are given by $\tau_\mu$ and $h_{\mu\nu}$ whereas the Carrollian metric-like objects are given by $v^\mu$ and $h_{\mu\nu}$ suggesting the duality \cite{Duval:2014uoa}
\begin{equation}\label{eq:metricity}
\tau_\mu\leftrightarrow v^\mu\,,\qquad h_{\mu\nu}\leftrightarrow h^{\mu\nu}\,,
\end{equation}
where TNC variables are written on the left and Carrollian fields on the right. When including the vector $M_\mu=m_\mu-\partial_\mu\chi$ for TNC geometry and $M^\mu$ for the Carrollian case we propose to extend this duality to
\begin{equation}\label{eq:boost}
M_\mu\leftrightarrow M^\mu\,,
\end{equation}
so that the remaining invariants are related by
\begin{equation}
\hat v^\mu\leftrightarrow\hat\tau_\mu\,,\qquad\bar h_{\mu\nu}\leftrightarrow\bar h^{\mu\nu}\,,\qquad\tilde\Phi\leftrightarrow\bar\Phi\,,
\end{equation}
where again on the left we have the TNC invariants $\bar h_{\mu\nu}$, $\hat v^\mu$ and $\tilde\Phi$ given in equations \eqref{eq:barh}, \eqref{eq:hatv} and \eqref{eq:tildePhi} (with $m_\mu$ replaced by $M_\mu$), respectively and on the right we have the Carrollian invariants $\bar h^{\mu\nu}$, $\hat \tau_\mu$ and $\bar\Phi$ given in equations \eqref{eq:hattau}, \eqref{eq:barinvh} and \eqref{eq:barPhi}, respectively. The duality \eqref{eq:metricity} and \eqref{eq:boost} interchanges the Galilean and Carrollian light cone structures in the sense that \eqref{eq:metricity} relates the notions of metricity while \eqref{eq:boost} swaps the notion of boost transformations. 

When there is no coupling to $\tilde\Phi$ on the TNC side and no coupling to $\bar\Phi$ on the Carrollian side, there is another relation between TNC and Carrollian geometry that interchanges like tensors as in \eqref{eq:swappingTNCandC}. For example if we apply this duality to the Carrollian affine connection \eqref{eq:manifestinvGamma2} which has the property that it does not depend on $\bar\Phi$ we obtain
\begin{equation}\label{eq:manifestinvGamma2TNC}
\Gamma^\lambda_{\mu\rho}=-\hat v^\lambda\partial_\mu\tau_\rho+\frac{1}{2}h^{\nu\lambda}\left(\partial_\mu\hat h_{\rho\nu}+\partial_\rho\hat h_{\mu\nu}-\partial_\nu\hat h_{\mu\rho}\right)-h^{\nu\lambda}\tau_\rho K_{\mu\nu}\,,
\end{equation}
where now the extrinsic curvature is given by $K_{\mu\nu}=-\tfrac{1}{2}\mathcal{L}_{\hat v}\hat h_{\mu\nu}$. We recognize the first two terms of \eqref{eq:manifestinvGamma2TNC} as the TNC connection that is independent of $\tilde\Phi$ used in \cite{Hartong:2015zia}. The third term containing the extrinsic curvature is just a harmless tensorial redefinition of the TNC connection. Put another way, in \cite{Hartong:2015zia} we used the connection 
\begin{equation}\label{eq:manifestinvGamma2TNC2}
\Gamma^\lambda_{\mu\rho}=-\hat v^\lambda\partial_\mu\tau_\rho+\frac{1}{2}h^{\nu\lambda}\left(\partial_\mu\hat h_{\rho\nu}+\partial_\rho\hat h_{\mu\nu}-\partial_\nu\hat h_{\mu\rho}\right)\,,
\end{equation}
obeying $\nabla_\mu \hat v^\nu=-h^{\nu\rho}K_{\nu\rho}$ but we could have equally absorbed the right hand side into the TNC connection leading to \eqref{eq:manifestinvGamma2TNC} which obeys $\nabla_\mu \hat v^\nu=0$. This direct relation between TNC and Carrollian affine connections does not extend to cases where the connections depend on $\tilde\Phi$ or $\bar\Phi$ as is obvious from the fact that then for example a Carrollian connection no longer has the property that $\nabla_\mu\hat\tau_\mu=0$ (see also the discussion at the end of section \ref{subsec:vectorM}).

\section{Ultra-Relativistic Gravity}\label{sec:Carrollgrav}

In \cite{Hartong:2015zia} it was shown how one can make TNC geometries dynamical by using an effective field theory approach where one writes all relevant and marginal terms that are second order in time derivatives, preserve time reversal invariance leading to the most general forms of Ho\v rava--Lifshitz gravity. Here we will start such an analysis for the case of dynamical Carrollian geometries. Since these have an ultra-relativistic light cone structure we will refer to the resulting theories as ultra-relativistic gravity. In order to decide whether a term is relevant, marginal or irrelevant we need to assign dilatation weights to the Carrollian fields $\tau_\mu$, $e_\mu^a$ and $M^\mu$.

We can extend the Carroll algebra by adding dilatations $D$ to it resulting in the Lifshitz--Carroll algebra\footnote{This algebra with $z=0$ is realized in higher dimensional uplifts of Lifshitz space-times. For example a $z=2$ Lifshitz space-time can be uplifted to a 5-dimensional $z=0$ Schr\"odinger space-time \cite{Balasubramanian:2010uk,Donos:2010tu,Costa:2010cn,Cassani:2011sv,Chemissany:2011mb}. In order to support this geometry one needs to add an axionic scalar which breaks the $z=0$ Schr\"odinger algebra down to the $z=0$ Lifshitz--Carroll algebra.} \cite{Gibbons:2009me,Bergshoeff:2015wma} whose extra commutators involving $D$ are
\begin{equation}
\left[D,H\right]=-zH\,,\qquad \left[D,P_a\right]=-P_a\,,\qquad\left[D,C_a\right]=(1-z)C_a\,.
\end{equation}

We can thus assign dilatation weight $-z$ to $\tau_\mu$ and $-1$ to $e_\mu^a$. Further in order that $\hat\tau_\mu$ and $\tau_\mu$ have the same dilatation weights we assign a weight $2-z$ to $M^\mu$, i.e. under a local $D$ transformation with parameter $\Lambda_D$ we have
\begin{eqnarray}
\delta_D \tau_\mu & = & z\Lambda_D\tau_\mu\,,\\
\delta_D e_\mu^a & = & \Lambda_D e_\mu^a\,,\\
\delta_D M^\mu & = & -(2-z)\Lambda_D M^\mu\,,
\end{eqnarray}
so that $\bar\Phi$ has dilatation weight $2(1-z)$, i.e.
\begin{equation}
\delta_D\bar\Phi=-2(1-z)\Lambda_D\bar\Phi\,.
\end{equation}
Note that $\tau_\mu$ and $e_\mu^a$ have the same dilatation weights as in the case of TNC geometry but that the weight of $\bar\Phi$ is opposite to that of $\tilde\Phi$. The reason for this is that in TNC geometry the vector $M_\mu$ has dilatation weight $z-2$ as follows for example from demanding that $\hat v^\mu$ and $v^\mu$ in \eqref{eq:hatv} both have the same dilatation weight $z$.

We will next consider actions in 2+1 dimensions with $0\le z<1$ by demanding local Carrollian invariance, i.e. by demanding that the Carrollian fields $\tau_\mu$, $e_\mu^a$ and $M^\mu$ only enter the action via the invariants $\hat\tau_\mu$, $h_{\mu\nu}$ and $\bar\Phi$. Further we will impose that the action is at most second order in time derivatives and preserves time reversal invariance. 

It is instructive to first consider the case with no coupling to $\bar\Phi$. As can be expected from the observations of section \ref{subsec:comparing}, where it is shown that a Carrollian geometry without $\bar\Phi$ can be obtained from a TNC geometry without $\tilde\Phi$ by interchanging like tensors as in \eqref{eq:swappingTNCandC}, the resulting actions should be of the HL form, but with $0\le z<1$. Indeed using the results of \cite{Hartong:2015zia} and the map \eqref{eq:swappingTNCandC} the following action is consistent with our coupling prescription for Carrollian gravity in $2+1$ dimensions with $0\le z<1$
\begin{equation}\label{eq:CarrHLaction}
S=\int d^3x e\left[C\left(K_{\mu\nu}K_{\rho\sigma}\hat h^{\mu\rho}\hat h^{\nu\sigma}-\lambda\left(\hat h^{\mu\nu}K_{\mu\nu}\right)^2\right)-\mathcal{V}\right]\,,
\end{equation}
where $e=\text{det}\,(\tau_\mu\,, e_\mu^a)$ which is invariant under local $C$ and $J$ transformations, where $K_{\mu\nu}=-\tfrac{1}{2}\mathcal{L}_v h_{\mu\nu}$ is the extrinsic curvature and where the potential $\mathcal{V}$ is taken to be
\begin{equation}
\mathcal{V}=-2\Lambda+c_1\mathcal{R}+c_2\hat h^{\mu\nu} a_\mu a_\nu \,.
\end{equation}
In here we defined 
\begin{eqnarray}
\mathcal{R} & = & \hat e^\mu_a \hat e^\nu_b R_{\mu\nu}{}^{ab}(J)=-\hat e^\mu_a \hat e^\nu_b \hat e^{\sigma a} e_\rho^bR_{\mu\nu\sigma}{}^\rho\,,\\
a_\mu & = & \mathcal{L}_v\hat\tau_\mu\,,
\end{eqnarray}
where the Riemann tensor $R_{\mu\rho\nu}{}^\rho$ is defined in \eqref{eq:Riemann} with the connection \eqref{eq:manifestinvGamma2} and where $a_\mu=\mathcal{L}_v\hat\tau_\mu$ is the Carrollian counterpart of the TNC torsion vector $a_\mu=\mathcal{L}_{\hat v}\tau_\mu$ \cite{Hartong:2015zia}. An action of this type with $\lambda=1$ and no potential term was considered in \cite{Teitelboim:1981fb,Henneaux:1979vn} as resulting from the $c\rightarrow 0$ limit of the Einstein--Hilbert action\footnote{I would like to thank Marc Henneaux for useful discussions about the $c\rightarrow 0$ limit of the Einstein--Hilbert action.}. All terms in \eqref{eq:CarrHLaction} are relevant for $0<z<1$ because the potential apart from the cosmological constant term involves terms of dilatation weight $2$ and the kinetic terms have dilatation weight $2z$ all of which are less than $2+z$ which is the negative of the dilatation weight of the integration measure $e$. The case with $z=0$ will be studied separately below. The dimensionless parameter $\lambda$ is the same as the one appearing in HL gravity \cite{Horava:2008ih,Horava:2009uw}.

Let us now introduce the scalar $\bar\Phi$. The first thing to observe is that for any $z\ge 0$ we can add the following coupling to the kinetic terms
\begin{equation}
\bar\Phi\left(K_{\mu\nu}K_{\rho\sigma}\hat h^{\mu\rho}\hat h^{\nu\sigma}-\lambda\left(\hat h^{\mu\nu}K_{\mu\nu}\right)^2\right)\,,
\end{equation}
since this has dilatation weight $2$ which is less than $z+2$. Further we can always add a term linear in $\bar\Phi$ to the potential since $2(1-z)\le 2+z$ for $0\le z<1$. On the other hand couplings such as $\bar\Phi\mathcal{R}$ or a kinetic term for $\bar\Phi$ such as $\left(v^\mu\partial_\mu\bar\Phi\right)^2$ have dilatation weight $4-2z$ and so in order that this is less than $z+2$ we need $z>2/3$. We will not consider such terms as we are primarily interested in those terms that are generic for $0\le z<1$. When we include $\bar\Phi$ we are thus led to the more general action
\begin{equation}\label{eq:CarrHLaction2}
S=\int d^3x e\left[\left(C+C_1\bar\Phi\right)\left(K_{\mu\nu}K_{\rho\sigma}\hat h^{\mu\rho}\hat h^{\nu\sigma}-\lambda\left(\hat h^{\mu\nu}K_{\mu\nu}\right)^2\right)-\mathcal{V}\right]\,,
\end{equation}
where the potential is given by
\begin{equation}
\mathcal{V}=-2\Lambda+c_1\mathcal{R}+c_2\hat h^{\mu\nu} a_\mu a_\nu +c_3\bar\Phi\,.
\end{equation}
The equation of motion of $\bar\Phi$ imposes the constraint 
\begin{equation}
K_{\mu\nu}K_{\rho\sigma}\hat h^{\mu\rho}\hat h^{\nu\sigma}-\lambda\left(\hat h^{\mu\nu}K_{\mu\nu}\right)^2=\frac{c_3}{C_1}\,.
\end{equation}
On the other hand the variation with respect to $h_{\mu\nu}$ will bring time derivatives onto $\bar\Phi$ upon partial integration making the scalar $\bar\Phi$ dynamical. It is interesting to contrast this with the case $1<z\le 2$ where we couple to TTNC geometry in the presence of $\tilde\Phi$ (section 11 of \cite{Hartong:2015zia}) where the field $\tilde\Phi$ imposes constraints on the terms in the potential rather than on the kinetic terms. The parameters in \eqref{eq:CarrHLaction2} have the following mass dimensions
\begin{equation}
[C]=M^{2-z}\,,\quad [C_1]=M^z\,,\quad [\Lambda]=M^{2+z}\,,\quad [c_1]=[c_2]=M^z\,,\quad [c_3]=M^{3z}\,.
\end{equation}

Finally we consider the special case $z=0$ and show that one can construct a local dilatation invariant action, i.e. an action with anisotropic Weyl invariance. Using that for $z=0$ the integration measure $e$ has weight $-2$ we need to construct terms with weight $2$. Under local dilatations the extrinsic curvature transforms as (for general $z$)
\begin{equation}
\delta_D K_{\mu\nu}=(2-z)\Lambda_D K_{\mu\nu}-h_{\mu\nu}v^\rho\partial_\rho\Lambda_D\,.
\end{equation}
It follows that $K_{\mu\nu}K_{\rho\sigma}\hat h^{\mu\rho}\hat h^{\nu\sigma}-\frac{1}{2}\left(\hat h^{\mu\nu}K_{\mu\nu}\right)^2$ is invariant under local scale transformations with weight $2z$. Using that for $z=0$ the scalar $\bar\Phi$ has weight 2 we find that the following term
\begin{equation}
\bar\Phi\left(K_{\mu\nu}K_{\rho\sigma}\hat h^{\mu\rho}\hat h^{\nu\sigma}-\frac{1}{2}\left(\hat h^{\mu\nu}K_{\mu\nu}\right)^2\right)\,,
\end{equation}
has weight 2 for $z=0$. Other terms with weight 2 are
\begin{align}
&\hat h^{\mu\nu}a_\mu a_\nu\,,\nonumber\\
&\bar\Phi\,,\\
&\hat h^{\mu\nu}a_\mu a_\nu\left(K_{\mu\nu}K_{\rho\sigma}\hat h^{\mu\rho}\hat h^{\nu\sigma}-\frac{1}{2}\left(\hat h^{\mu\nu}K_{\mu\nu}\right)^2\right)\,.\nonumber
\end{align}
Hence the following action has anisotropic Weyl invariance with $z=0$
\begin{equation}\label{eq:ConfCarrHLaction}
S=\int d^3x e\left[\left(C_1\bar\Phi+C_2\hat h^{\mu\nu} a_\mu a_\nu\right)\left(K_{\mu\nu}K_{\rho\sigma}\hat h^{\mu\rho}\hat h^{\nu\sigma}-\frac{1}{2}\left(\hat h^{\mu\nu}K_{\mu\nu}\right)^2\right)-\mathcal{V}\right]\,,
\end{equation}
where the potential is given by
\begin{equation}
\mathcal{V}=c_2\hat h^{\mu\nu} a_\mu a_\nu +c_3\bar\Phi\,.
\end{equation}
This action with anisotropic Weyl invariance for $z=0$ only has dimensionless coupling constants. 

We note that the spatial Ricci scalar $\mathcal{R}$ transforms under local $D$ transformations as (in $d=2$ spatial dimensions)
\begin{equation}
\delta_D\mathcal{R}=-2\Lambda_D\mathcal{R}+2\hat h^{\mu\nu}\nabla_\mu\partial_\nu\Lambda_D\,.
\end{equation}
Different from the conformal TNC case (section 12 of \cite{Hartong:2015zia}) here we cannot use the vector $a_\mu$ to make a local $D$ invariant combination out of $\mathcal{R}$ and derivatives of $a_\mu$ because for $z=0$ the vector $a_\mu$ is invariant under local $D$ transformations.

%

\section{Discussion}

It would be interesting to extend this work in the following directions. 

It has been known for a long time that the asymptotic symmetry algebra of asymptotically flat space-times is given by the Bondi--Metzner--Sachs (BMS) algebra \cite{Bondi:1962px,Sachs:1962wk,Sachs:1962zza} (see also \cite{Barnich:2009se,Barnich:2010eb}). In 3 bulk dimensions it has been shown that the BMS algebra is isomorphic to the 2-dimensional Galilean conformal algebra \cite{Bagchi:2010zz,Bagchi:2012cy} (which is a contraction of the relativistic conformal group \cite{Bagchi:2009my}). Recently conformal extensions of the Carroll algebra have been studied in \cite{Duval:2014lpa,Duval:2014uva} and it has been shown that the BMS algebra forms a conformal extension of the Carroll algebra \cite{Duval:2014uva}. Regarding the case of flat space holography in 3 bulk dimensions the Galilean structures seen at infinity can be interpreted as Carrollian because in 1+1 boundary dimensions interchanging space and time leads to an isomorphism between the Carroll and Galilei algebras. Further, future and past null infinity form Carrollian space-times \cite{Duval:2014uva}. It could therefore be insightful to explore the connections between the gauging of the Carroll algebra and flat space holography further.

The space-time symmetries of warped conformal field theories involve Carrollian boosts that together with the scale transformations form the $z=0$ Lifshitz--Carroll algebra \cite{Hofman:2014loa}. It would be interesting to apply the methods for the gauging of the Carroll algebra as performed here to study the coupling of these WCFTs to curved backgrounds.  

More generally along similar lines one can couple field theories to Carrollian geometries and study global symmetries by defining conformal Killing vectors, define an energy-momentum tensor by varying the invariants $\hat \tau_\mu$ and $h_{\mu\nu}$ much like it was done for field theories coupled to TNC geometries \cite{Geracie:2014nka,Jensen:2014aia,Hartong:2014pma,Hartong:2015wxa}. It would be interesting to understand what the role of the scalar $\bar\Phi$ is when coupling field theories to Carrollian geometries, i.e. to understand what the response is to varying this background field.

Finally, one can study the actions for ultra-relativistic or Carrollian gravity further by e.g. studying their phase space formulation, count the number of degrees of freedom, etc. It would be interesting to generalize the 3-dimensional actions of ultra-relativistic gravity constructed here to higher dimensions and to study the equations of motion by looking for various classes of solutions such as cosmological and spherically symmetric space-times. It would be interesting to study the perturbative properties of these theories for example by linearizing around flat space-time and study the form of the propagators, etc.

\section*{Acknowledgments}

I would like to thank Niels Obers for valuable discussions and careful reading of this manuscript. Further I would like to thank Diego Hofman and Blaise Rollier for the very insightful discussions regarding their work \cite{Hofman:2014loa}. The work of JH is supported by the advanced ERC grant `Symmetries and Dualities in Gravity and M-theory' of Marc Henneaux. I thank the Galileo Galilei Institute for Theoretical Physics for the hospitality and the INFN for partial support during the initial stages of this work.

%
%
%

\begin{thebibliography}{10}

\bibitem{Hartong:2014oma}
J.~Hartong, E.~Kiritsis, and N.~A. Obers, ``{Lifshitz Space-Times for
  Schroedinger Holography},''
\href{http://arxiv.org/abs/1409.1519}{{\tt arXiv:1409.1519 [hep-th]}}.

\bibitem{Bergshoeff:2014uea}
E.~A. Bergshoeff, J.~Hartong, and J.~Rosseel, ``{Torsional Newton-Cartan
  Geometry and the Schr\"odinger Algebra},''
\href{http://arxiv.org/abs/1409.5555}{{\tt arXiv:1409.5555 [hep-th]}}.

\bibitem{Hartong:2015wxa}
J.~Hartong, E.~Kiritsis, and N.~A. Obers, ``{Field Theory on Newton-Cartan
  Backgrounds and Symmetries of the Lifshitz Vacuum},''
\href{http://arxiv.org/abs/1502.00228}{{\tt arXiv:1502.00228 [hep-th]}}.

\bibitem{Guica:2010sw}
M.~Guica, K.~Skenderis, M.~Taylor, and B.~C. van Rees, ``{Holography for
  Schrodinger backgrounds},''
  \href{http://dx.doi.org/10.1007/JHEP02(2011)056}{{\em JHEP} {\bf 1102} (2011)
   056},
\href{http://arxiv.org/abs/1008.1991}{{\tt arXiv:1008.1991 [hep-th]}}.

\bibitem{Hartong:2010ec}
J.~Hartong and B.~Rollier, ``{Asymptotically Schroedinger Space-Times: TsT
  Transformations and Thermodynamics},''
  \href{http://dx.doi.org/10.1007/JHEP01(2011)084}{{\em JHEP} {\bf 1101} (2011)
   084},
\href{http://arxiv.org/abs/1009.4997}{{\tt arXiv:1009.4997 [hep-th]}}.

\bibitem{Hartong:2013cba}
J.~Hartong and B.~Rollier, ``{Particle Number and 3D Schršdinger Holography},''
  \href{http://dx.doi.org/10.1007/JHEP09(2014)111}{{\em JHEP} {\bf 1409} (2014)
   111},
\href{http://arxiv.org/abs/1305.3653}{{\tt arXiv:1305.3653 [hep-th]}}.

\bibitem{Andrade:2014iia}
T.~Andrade, C.~Keeler, A.~Peach, and S.~F. Ross, ``{Schrodinger Holography for
  $z<2$},''
\href{http://arxiv.org/abs/1408.7103}{{\tt arXiv:1408.7103 [hep-th]}}.

\bibitem{Andrade:2014kba}
T.~Andrade, C.~Keeler, A.~Peach, and S.~F. Ross, ``{Schr\"odinger Holography
  with $z=2$},''
\href{http://arxiv.org/abs/1412.0031}{{\tt arXiv:1412.0031 [hep-th]}}.

\bibitem{deHaro:2000xn}
S.~de~Haro, S.~N. Solodukhin, and K.~Skenderis, ``{Holographic reconstruction
  of space-time and renormalization in the AdS / CFT correspondence},''
  \href{http://dx.doi.org/10.1007/s002200100381}{{\em Commun.Math.Phys.} {\bf
  217} (2001)  595--622},
\href{http://arxiv.org/abs/hep-th/0002230}{{\tt arXiv:hep-th/0002230
  [hep-th]}}.

\bibitem{Papadimitriou:2005ii}
I.~Papadimitriou and K.~Skenderis, ``{Thermodynamics of asymptotically locally
  AdS spacetimes},''
  \href{http://dx.doi.org/10.1088/1126-6708/2005/08/004}{{\em JHEP} {\bf 0508}
  (2005)  004},
\href{http://arxiv.org/abs/hep-th/0505190}{{\tt arXiv:hep-th/0505190
  [hep-th]}}.

\bibitem{LŽvy1965}
J.-M. Levy-Leblond, ``Une nouvelle limite non-relativiste du groupe de
  poincar\'e,'' {\em Annales de l'institut Henri Poincar\'e (A) Physique
  th\'eorique} {\bf 3} (1965) no.~1, 1--12. \url{http://eudml.org/doc/75509}.

\bibitem{Bacry:1968zf}
H.~Bacry and J.~Levy-Leblond, ``{Possible kinematics},''
\href{http://dx.doi.org/10.1063/1.1664490}{{\em J.Math.Phys.} {\bf 9} (1968)
  1605--1614}.

\bibitem{Duval:2014uva}
C.~Duval, G.~Gibbons, and P.~Horvathy, ``{Conformal Carroll groups and BMS
  symmetry},'' \href{http://dx.doi.org/10.1088/0264-9381/31/9/092001}{{\em
  Class.Quant.Grav.} {\bf 31} (2014)  092001},
\href{http://arxiv.org/abs/1402.5894}{{\tt arXiv:1402.5894 [gr-qc]}}.

\bibitem{Hofman:2014loa}
D.~M. Hofman and B.~Rollier, ``{Warped Conformal Field Theory as Lower Spin
  Gravity},''
\href{http://arxiv.org/abs/1411.0672}{{\tt arXiv:1411.0672 [hep-th]}}.

\bibitem{Hartong:2015zia}
J.~Hartong and N.~A. Obers, ``{Horava-Lifshitz Gravity From Dynamical
  Newton-Cartan Geometry},''
\href{http://arxiv.org/abs/1504.07461}{{\tt arXiv:1504.07461 [hep-th]}}.

\bibitem{Christensen:2013lma}
M.~H. Christensen, J.~Hartong, N.~A. Obers, and B.~Rollier, ``{Torsional
  Newton-Cartan Geometry and Lifshitz Holography},''
  \href{http://dx.doi.org/10.1103/PhysRevD.89.061901}{{\em Phys.Rev.} {\bf D89}
  (2014)  061901},
\href{http://arxiv.org/abs/1311.4794}{{\tt arXiv:1311.4794 [hep-th]}}.

\bibitem{Christensen:2013rfa}
M.~H. Christensen, J.~Hartong, N.~A. Obers, and B.~Rollier, ``{Boundary
  Stress-Energy Tensor and Newton-Cartan Geometry in Lifshitz Holography},''
  \href{http://dx.doi.org/10.1007/JHEP01(2014)057}{{\em JHEP} {\bf 1401} (2014)
   057},
\href{http://arxiv.org/abs/1311.6471}{{\tt arXiv:1311.6471 [hep-th]}}.

\bibitem{Son:2013rqa}
D.~T. Son, ``{Newton-Cartan Geometry and the Quantum Hall Effect},''
\href{http://arxiv.org/abs/1306.0638}{{\tt arXiv:1306.0638
  [cond-mat.mes-hall]}}.

\bibitem{Geracie:2014nka}
M.~Geracie, D.~T. Son, C.~Wu, and S.-F. Wu, ``{Spacetime Symmetries of the
  Quantum Hall Effect},''
  \href{http://dx.doi.org/10.1103/PhysRevD.91.045030}{{\em Phys.Rev.} {\bf D91}
  (2015)  045030},
\href{http://arxiv.org/abs/1407.1252}{{\tt arXiv:1407.1252
  [cond-mat.mes-hall]}}.

\bibitem{Banerjee:2014nja}
R.~Banerjee, A.~Mitra, and P.~Mukherjee, ``{Localization of the Galilean
  symmetry and dynamical realization of Newton-Cartan geometry},''
  \href{http://dx.doi.org/10.1088/0264-9381/32/4/045010}{{\em
  Class.Quant.Grav.} {\bf 32} (2015) no.~4, 045010},
\href{http://arxiv.org/abs/1407.3617}{{\tt arXiv:1407.3617 [hep-th]}}.

\bibitem{Brauner:2014jaa}
T.~Brauner, S.~Endlich, A.~Monin, and R.~Penco, ``{General coordinate
  invariance in quantum many-body systems},''
  \href{http://dx.doi.org/10.1103/PhysRevD.90.105016}{{\em Phys.Rev.} {\bf D90}
  (2014) no.~10, 105016},
\href{http://arxiv.org/abs/1407.7730}{{\tt arXiv:1407.7730 [hep-th]}}.

\bibitem{Jensen:2014aia}
K.~Jensen, ``{On the coupling of Galilean-invariant field theories to curved
  spacetime},''
\href{http://arxiv.org/abs/1408.6855}{{\tt arXiv:1408.6855 [hep-th]}}.

\bibitem{Hartong:2014pma}
J.~Hartong, E.~Kiritsis, and N.~A. Obers, ``{Schroedinger Invariance from
  Lifshitz Isometries in Holography and Field Theory},''
\href{http://arxiv.org/abs/1409.1522}{{\tt arXiv:1409.1522 [hep-th]}}.

\bibitem{Bekaert:2014bwa}
X.~Bekaert and K.~Morand, ``{Connections and dynamical trajectories in
  generalised Newton-Cartan gravity I. An intrinsic view},''
\href{http://arxiv.org/abs/1412.8212}{{\tt arXiv:1412.8212 [hep-th]}}.

\bibitem{Andringa:2010it}
R.~Andringa, E.~Bergshoeff, S.~Panda, and M.~de~Roo, ``{Newtonian Gravity and
  the Bargmann Algebra},''
  \href{http://dx.doi.org/10.1088/0264-9381/28/10/105011}{{\em
  Class.Quant.Grav.} {\bf 28} (2011)  105011},
\href{http://arxiv.org/abs/1011.1145}{{\tt arXiv:1011.1145 [hep-th]}}.

\bibitem{Duval:2014uoa}
C.~Duval, G.~Gibbons, P.~Horvathy, and P.~Zhang, ``{Carroll versus Newton and
  Galilei: two dual non-Einsteinian concepts of time},''
  \href{http://dx.doi.org/10.1088/0264-9381/31/8/085016}{{\em
  Class.Quant.Grav.} {\bf 31} (2014)  085016},
\href{http://arxiv.org/abs/1402.0657}{{\tt arXiv:1402.0657 [gr-qc]}}.

\bibitem{Horava:2008ih}
P.~Horava, ``{Membranes at Quantum Criticality},''
  \href{http://dx.doi.org/10.1088/1126-6708/2009/03/020}{{\em JHEP} {\bf 0903}
  (2009)  020},
\href{http://arxiv.org/abs/0812.4287}{{\tt arXiv:0812.4287 [hep-th]}}.

\bibitem{Horava:2009uw}
P.~Horava, ``{Quantum Gravity at a Lifshitz Point},''
  \href{http://dx.doi.org/10.1103/PhysRevD.79.084008}{{\em Phys.Rev.} {\bf D79}
  (2009)  084008},
\href{http://arxiv.org/abs/0901.3775}{{\tt arXiv:0901.3775 [hep-th]}}.

\bibitem{Horava:2010zj}
P.~Horava and C.~M. Melby-Thompson, ``{General Covariance in Quantum Gravity at
  a Lifshitz Point},'' \href{http://dx.doi.org/10.1103/PhysRevD.82.064027}{{\em
  Phys.Rev.} {\bf D82} (2010)  064027},
\href{http://arxiv.org/abs/1007.2410}{{\tt arXiv:1007.2410 [hep-th]}}.

\bibitem{Teitelboim:1981fb}
C.~Teitelboim, ``{The Hamiltonian Structure of Space-Time},''
{\em General Relativity and Gravitation, Vol.1, 195-225, Ed. Held, A.} (1981)
  .

\bibitem{Henneaux:1979vn}
M.~Henneaux, ``{Geometry of Zero Signature Space-times},''
{\em Bull.Soc.Math.Belg.} {\bf 31} (1979)  47--63.

\bibitem{Dautcourt:1997hb}
G.~Dautcourt, ``{On the ultrarelativistic limit of general relativity},'' {\em
  Acta Phys.Polon.} {\bf B29} (1998)  1047--1055,
\href{http://arxiv.org/abs/gr-qc/9801093}{{\tt arXiv:gr-qc/9801093 [gr-qc]}}.

\bibitem{Anderson:2002zn}
E.~Anderson, ``{Strong coupled relativity without relativity},''
  \href{http://dx.doi.org/10.1023/B:GERG.0000010474.63835.2c}{{\em
  Gen.Rel.Grav.} {\bf 36} (2004)  255--276},
\href{http://arxiv.org/abs/gr-qc/0205118}{{\tt arXiv:gr-qc/0205118 [gr-qc]}}.

\bibitem{Gibbons:2002tv}
G.~Gibbons, K.~Hashimoto, and P.~Yi, ``{Tachyon condensates, Carrollian
  contraction of Lorentz group, and fundamental strings},''
  \href{http://dx.doi.org/10.1088/1126-6708/2002/09/061}{{\em JHEP} {\bf 0209}
  (2002)  061},
\href{http://arxiv.org/abs/hep-th/0209034}{{\tt arXiv:hep-th/0209034
  [hep-th]}}.

\bibitem{Damour:2002et}
T.~Damour, M.~Henneaux, and H.~Nicolai, ``{Cosmological billiards},''
  \href{http://dx.doi.org/10.1088/0264-9381/20/9/201}{{\em Class.Quant.Grav.}
  {\bf 20} (2003)  R145--R200},
\href{http://arxiv.org/abs/hep-th/0212256}{{\tt arXiv:hep-th/0212256
  [hep-th]}}.

\bibitem{Bekaert:2015xua}
X.~Bekaert and K.~Morand, ``{Connections and dynamical trajectories in
  generalised Newton-Cartan gravity II. An ambient perspective},''
\href{http://arxiv.org/abs/1505.03739}{{\tt arXiv:1505.03739 [hep-th]}}.

\bibitem{Bergshoeff:2014jla}
E.~Bergshoeff, J.~Gomis, and G.~Longhi, ``{Dynamics of Carroll Particles},''
  \href{http://dx.doi.org/10.1088/0264-9381/31/20/205009}{{\em
  Class.Quant.Grav.} {\bf 31} (2014) no.~20, 205009},
\href{http://arxiv.org/abs/1405.2264}{{\tt arXiv:1405.2264 [hep-th]}}.

\bibitem{Ortin:2004ms}
T.~Ortin, ``{Gravity and strings},''
{\em Cambridge Unversity, Cambridge University Press} (2004)  .

\bibitem{Duval:1984cj}
C.~Duval, G.~Burdet, H.~Kunzle, and M.~Perrin, ``{Bargmann Structures and
  Newton-Cartan Theory},''
\href{http://dx.doi.org/10.1103/PhysRevD.31.1841}{{\em Phys.Rev.} {\bf D31}
  (1985)  1841}.

\bibitem{Julia:1994bs}
B.~Julia and H.~Nicolai, ``{Null Killing vector dimensional reduction and
  Galilean geometrodynamics},''
  \href{http://dx.doi.org/10.1016/0550-3213(94)00584-2}{{\em Nucl.Phys.} {\bf
  B439} (1995)  291--326},
\href{http://arxiv.org/abs/hep-th/9412002}{{\tt arXiv:hep-th/9412002
  [hep-th]}}.

\bibitem{Bekaert:2013fta}
X.~Bekaert and K.~Morand, ``{Embedding non-relativistic physics inside a
  gravitational wave},''
\href{http://arxiv.org/abs/1307.6263}{{\tt arXiv:1307.6263 [hep-th]}}.

\bibitem{Balasubramanian:2010uk}
K.~Balasubramanian and K.~Narayan, ``{Lifshitz spacetimes from AdS null and
  cosmological solutions},''
  \href{http://dx.doi.org/10.1007/JHEP08(2010)014}{{\em JHEP} {\bf 1008} (2010)
   014},
\href{http://arxiv.org/abs/1005.3291}{{\tt arXiv:1005.3291 [hep-th]}}.

\bibitem{Donos:2010tu}
A.~Donos and J.~P. Gauntlett, ``{Lifshitz Solutions of D=10 and D=11
  supergravity},'' \href{http://dx.doi.org/10.1007/JHEP12(2010)002}{{\em JHEP}
  {\bf 1012} (2010)  002},
\href{http://arxiv.org/abs/1008.2062}{{\tt arXiv:1008.2062 [hep-th]}}.

\bibitem{Costa:2010cn}
R.~Caldeira~Costa and M.~Taylor, ``{Holography for chiral scale-invariant
  models},'' \href{http://dx.doi.org/10.1007/JHEP02(2011)082}{{\em JHEP} {\bf
  1102} (2011)  082},
\href{http://arxiv.org/abs/1010.4800}{{\tt arXiv:1010.4800 [hep-th]}}.

\bibitem{Cassani:2011sv}
D.~Cassani and A.~F. Faedo, ``{Constructing Lifshitz solutions from AdS},''
  \href{http://dx.doi.org/10.1007/JHEP05(2011)013}{{\em JHEP} {\bf 1105} (2011)
   013},
\href{http://arxiv.org/abs/1102.5344}{{\tt arXiv:1102.5344 [hep-th]}}.

\bibitem{Chemissany:2011mb}
W.~Chemissany and J.~Hartong, ``{From D3-Branes to Lifshitz Space-Times},''
  \href{http://dx.doi.org/10.1088/0264-9381/28/19/195011}{{\em
  Class.Quant.Grav.} {\bf 28} (2011)  195011},
\href{http://arxiv.org/abs/1105.0612}{{\tt arXiv:1105.0612 [hep-th]}}.

\bibitem{Gibbons:2009me}
G.~Gibbons, J.~Gomis, and C.~Pope, ``{Deforming the Maxwell-Sim Algebra},''
  \href{http://dx.doi.org/10.1103/PhysRevD.82.065002}{{\em Phys.Rev.} {\bf D82}
  (2010)  065002},
\href{http://arxiv.org/abs/0910.3220}{{\tt arXiv:0910.3220 [hep-th]}}.

\bibitem{Bergshoeff:2015wma}
E.~Bergshoeff, J.~Gomis, and L.~Parra, ``{The Symmetries of the Carroll
  Superparticle},''
\href{http://arxiv.org/abs/1503.06083}{{\tt arXiv:1503.06083 [hep-th]}}.

\bibitem{Bondi:1962px}
H.~Bondi, M.~van~der Burg, and A.~Metzner, ``{Gravitational waves in general
  relativity. 7. Waves from axisymmetric isolated systems},''
\href{http://dx.doi.org/10.1098/rspa.1962.0161}{{\em Proc.Roy.Soc.Lond.} {\bf
  A269} (1962)  21--52}.

\bibitem{Sachs:1962wk}
R.~Sachs, ``{Gravitational waves in general relativity. 8. Waves in
  asymptotically flat space-times},''
\href{http://dx.doi.org/10.1098/rspa.1962.0206}{{\em Proc.Roy.Soc.Lond.} {\bf
  A270} (1962)  103--126}.

\bibitem{Sachs:1962zza}
R.~Sachs, ``{Asymptotic symmetries in gravitational theory},''
\href{http://dx.doi.org/10.1103/PhysRev.128.2851}{{\em Phys.Rev.} {\bf 128}
  (1962)  2851--2864}.

\bibitem{Barnich:2009se}
G.~Barnich and C.~Troessaert, ``{Symmetries of asymptotically flat 4
  dimensional spacetimes at null infinity revisited},''
  \href{http://dx.doi.org/10.1103/PhysRevLett.105.111103}{{\em Phys.Rev.Lett.}
  {\bf 105} (2010)  111103},
\href{http://arxiv.org/abs/0909.2617}{{\tt arXiv:0909.2617 [gr-qc]}}.

\bibitem{Barnich:2010eb}
G.~Barnich and C.~Troessaert, ``{Aspects of the BMS/CFT correspondence},''
  \href{http://dx.doi.org/10.1007/JHEP05(2010)062}{{\em JHEP} {\bf 1005} (2010)
   062},
\href{http://arxiv.org/abs/1001.1541}{{\tt arXiv:1001.1541 [hep-th]}}.

\bibitem{Bagchi:2010zz}
A.~Bagchi, ``{Correspondence between Asymptotically Flat Spacetimes and
  Nonrelativistic Conformal Field Theories},''
\href{http://dx.doi.org/10.1103/PhysRevLett.105.171601}{{\em Phys.Rev.Lett.}
  {\bf 105} (2010)  171601}.

\bibitem{Bagchi:2012cy}
A.~Bagchi and R.~Fareghbal, ``{BMS/GCA Redux: Towards Flatspace Holography from
  Non-Relativistic Symmetries},''
  \href{http://dx.doi.org/10.1007/JHEP10(2012)092}{{\em JHEP} {\bf 1210} (2012)
   092},
\href{http://arxiv.org/abs/1203.5795}{{\tt arXiv:1203.5795 [hep-th]}}.

\bibitem{Bagchi:2009my}
A.~Bagchi and R.~Gopakumar, ``{Galilean Conformal Algebras and AdS/CFT},''
  \href{http://dx.doi.org/10.1088/1126-6708/2009/07/037}{{\em JHEP} {\bf 0907}
  (2009)  037},
\href{http://arxiv.org/abs/0902.1385}{{\tt arXiv:0902.1385 [hep-th]}}.

\bibitem{Duval:2014lpa}
C.~Duval, G.~Gibbons, and P.~Horvathy, ``{Conformal Carroll groups},''
  \href{http://dx.doi.org/10.1088/1751-8113/47/33/335204}{{\em J.Phys.} {\bf
  A47} (2014)  335204},
\href{http://arxiv.org/abs/1403.4213}{{\tt arXiv:1403.4213 [hep-th]}}.

\end{thebibliography}

\providecommand{\href}[2]{#2}\begingroup\raggedright\endgroup

\end{document}